# Moving forward with multilayer metastructures - simple design of complex optics

**DAVID A. B. MILLER**

*Ginzton Laboratory, Stanford University, 348 Via Pueblo Mall, Stanford CA 94305 USA*
*dabm@stanford.edu*

**Abstract:** I propose a simple and progressive way of designing and fabricating complex multilayered optics, even as we scale to large numbers of layers. The approach relies on successive layers of two-by-two blocks in which light only flows in the forward direction. It exploits a proposed trapezoidal architecture that can also support self-configuration in programmable structures, with the desired linear function or matrix set up using input vectors that correspond to the matrix rows, without any iteration other than successive single-parameter power maximizations. A key required component is a layer of 2 by 2 beamsplitters. The approach can support functions corresponding to a broad range of banded diagonal matrices. Example applications include a camera that also measures phase gradients and the direct convolution of a line of inputs with a kernel based on wavelets.

## 1. Introduction

Metamaterials have opened a wide range of possibilities in optics [1,2]. They can embed remarkable optical complexity in thin layers. Like complex photonics more generally, there are many potential applications in optics itself, for sensing, communicating and processing light, and possibly also in information processing through sophisticated linear functions with zero energy dissipation [3–8]. Yet, we understand that many functions cannot be performed in just one such layer [9]. Using the impressive and growing power of inverse design and other techniques [10,11], we can design complex multilayered structures for more sophisticated functionality [12]. However, designs with large numbers of custom layers of precise thickness and alignment could be very difficult to make. Inverse design may also not scale well to larger design volumes because of difficulty in simulating waves in large complex structures [11]. We face, too, the challenge that there may be limited applications for fixed complex designs that solve just one specific problem. We would therefore like some layered approach that would enable scalable design for a broad range of problems, that would lead to structures that were relatively easy and inexpensive to fabricate, and that possibly even would allow programmability or even self-configuration to the problem of interest.

Here we introduce an architecture for multilayered structures that has (i) simple progressive design and/or programming even for complex problems, (ii) simplifies manufacture by using standardized layers and avoiding sensitivity to precise layer thicknesses, and (iii) can even self-configure to problems of interest. This architecture is particularly appropriate where the number of input channels or pixels is much greater than the number of layers we can fabricate, as is naturally the case for layered metastructures. It could have immediate implementation in planar silicon photonic circuits, where Mach-Zehnder interferometer (MZI) building blocks are physically generally much longer than they are wide, leading to long, narrow circuits. It poses one challenge and opportunity – devise a layered metastructure that functions as a 2-dimensional array of $2\times2$ beamsplitters – though we can certainly propose various ways of doing this, at least in principle.

I introduce the architecture in Section 2 and analyze the simplest, single-layer case in Section 3, and show simple examples. Section 4 analyzes multilayer meshes and Section 5 gives an example design and application for wavelet-based filtering. Approaches to manufacturable multilayer meshes are discussed in Section 6, some extensions are briefly discussed in Section 7, and features and challenges are discussed in Section 8 before drawing conclusions in Section 9.

## 2. Architectural concept

The concept of the approach is illustrated in Fig. 1. This "trapezoidal" architecture consists of a number $m$ of layers of $2 \times 2$ blocks forming an interconnected mesh. Such blocks can be constructed as Mach-Zehnder interferometers (MZIs) with two control phases [13] – a (differential) input phase $\Delta\phi$ and a (differential) internal phase $\Delta\theta$ as in Fig. 1(a) – but other approaches are also possible, such as a simple beamsplitters with an input phase delay [14,15]. We can use the simplified notation of Fig. 1(b) for any such blocks. We presume that we have "pixelated" the input optical field into a set of separate inputs that can be fed individually into the block inputs. Such pixelation could be accomplished by an array of lenslets, for example, or we could just regard ourselves as sampling the incoming field at the points or small areas corresponding to the mesh inputs.

The core of the architecture is a trapezoid of such blocks as in the example for $m = 4$ layers of blocks in Fig. 1(c), which can then be extended laterally by any number of successive additional diagonal lines of blocks. With $m$ such layers, the bottom of the core trapezoid would have $m$ blocks, expanding to $2m-1$ blocks at its top.

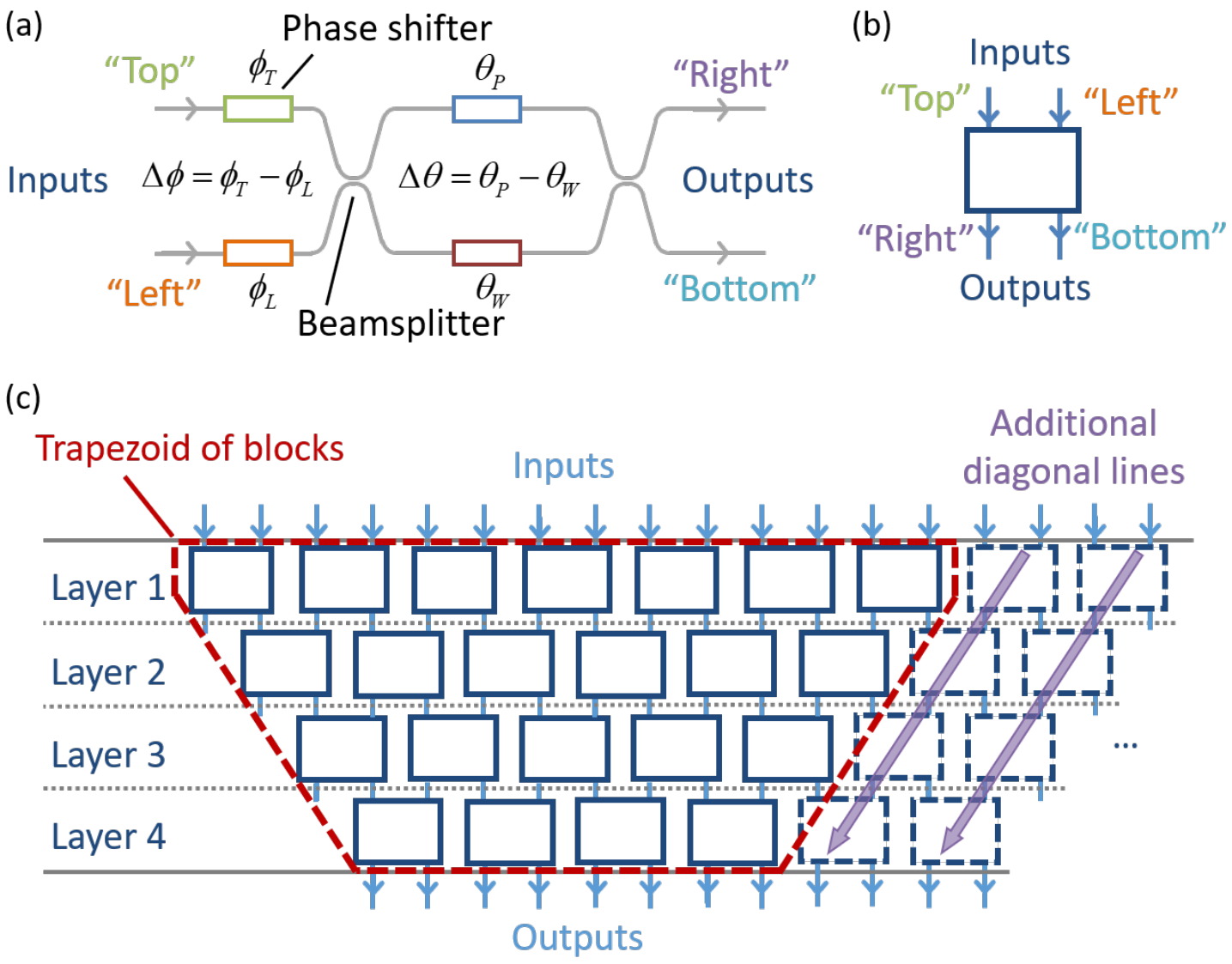

Fig. 1. Trapezoidal architecture for multilayered structure. (a) Conceptual illustration of a $2 \times 2$ block as a waveguide Mach-Zehnder interferometer controlled by two (differential) phase shifts, $\Delta\phi$ and $\Delta\theta$, with labeling of ports as in [13]. (Top, Left, Right and Bottom labels are by analogy with a cube beamsplitter [13]) (b) Simplified notation for a $2 \times 2$ block. (Note that the counterintuitive labelling of the ports here, especially "Left" and "Right", is required since we label Inputs and Outputs below from left to right in the figures.) (c) Explicit architecture as a forward-only mesh of $2 \times 2$ blocks, in an example with 4 layers. The architecture includes a core trapezoid of blocks that can then be extended laterally by additional diagonal lines of blocks to make a layered structure of arbitrary width.

A key point of this architecture is that light only flows in one direction, here from inputs on the top to outputs on the bottom. This "forward-only" aspect [14–16] is crucial in making such a structure easy to design, program or even self-configure [13–15,17–24]. ("Forward" means from top to bottom in Fig. 1.) It also avoids extreme sensitivity to layer thicknesses since there are no interferences forward and backwards in the structure.

One might think that restricting an architecture to being forward-only would also substantially restrict the functions it could perform; we already know, however, that forward-only architectures of sufficient depth can implement quite arbitrary linear transforms at a given wavelength [15]. Those architectures demonstrate that reflections back and forward inside a structure, which tend to happen implicitly in many layered structures, are not required for such arbitrary transforms. Reflections would be required to create spectral filters based on resonances; it has, however, recently been understood that forward-only systems can make arbitrary filters also [22,23] (at least in the finite impulse response case). Hence, the restriction of forward-only operation is arguably not a major constraint on what the device can do. Of course, especially with a limited number of layers, such a device is restricted to certain linear mappings, in particular to ones with a given overlapping nonlocality [9] (see also Supplemental document Section S1), but the same is true of any other approach to layered structures. We can also state quite explicitly the broad class of linear transforms that can be performed by our trapezoidal forward-only approach, and we return to this below.

Note, too, that, if the blocks are programmable, this architecture can be self-configured using a novel algorithm we present below in the sense that it can be physically set up by shining in just the physical vectors $|\psi_j\rangle$ of amplitudes in the set of inputs that correspond to the rows of the matrix that represents the device. (The matrix rows are formally the Hermitian adjoints $\langle\psi_j|$ of those input vectors.) If we shine in those vectors one by one, we can perform a progressive series of single-parameter power minimizations or maximizations as we adjust phase shifters (or, in non-MZI approaches, possibly also beamsplitter ratios) inside the blocks, also one by one. So, if the system has controllable phase shifters, it can be set up without calibration and can automatically compensate for minor path-length differences inside the system. The existence of such a physical configuration algorithm also means we can calibrate the system progressively [13], and we can mathematically progressively calculate the necessary settings of all the blocks for any desired design (even if the structure cannot be programmed after manufacture).

### 3. Analysis of single blocks and a single layer architecture

As a preliminary, we need to understand the function of an individual $2\times 2$ block. I analyze this in detail in Supplemental document Section S2, summarizing key results here. Presuming 50:50 beamsplitters, and neglecting the overall phase in passing through the MZI (any such common-mode phase shift $\phi_{CM}$ can be added as an overall multiplying factor $\exp(i\phi_{CM})$ later if needed), the matrix describing the block can be written [13] as

$$\mathsf{M}_T(\Delta\phi,\Delta\theta)=\begin{bmatrix}\exp\left(i\frac{\Delta\phi}{2}\right)\sin\left(\frac{\Delta\theta}{2}\right) & \exp\left(-i\frac{\Delta\phi}{2}\right)\cos\left(\frac{\Delta\theta}{2}\right)\\ \exp\left(i\frac{\Delta\phi}{2}\right)\cos\left(\frac{\Delta\theta}{2}\right) & -\exp\left(-i\frac{\Delta\phi}{2}\right)\sin\left(\frac{\Delta\theta}{2}\right)\end{bmatrix}\equiv\begin{bmatrix}u_{1,1} & u_{1,2}\\ u_{2,1} & u_{2,2}\end{bmatrix} \tag{1}$$

where we have also introduced a notation $u_{i,j}$ for the matrix element in the $i$th row and $j$th column. If write the input vector of mode amplitudes as

$$|e_{in}\rangle=\begin{bmatrix}a\\ b\end{bmatrix}\equiv\begin{bmatrix}|a|\exp(i\alpha)\\ |b|\exp(i\beta)\end{bmatrix} \tag{2}$$

with $a$ and $b$ being the electric field amplitudes in the "Top" and "Left" inputs respectively, the resulting output vector $|e_{out}\rangle$ of amplitudes $c$ and $d$ in the "Right" and "Bottom" outputs respectively is given by

$$|e_{out}\rangle \equiv \begin{bmatrix} c \\ d \end{bmatrix} = \mathsf{M}_T |e_{in}\rangle \tag{3}$$

For our later discussions, it is also useful to understand how to self-configure such a block to route all the power in an arbitrary pair of inputs to a given output [13–15,17,25]. Here, with any mutually coherent relative amplitudes and phases in the "Top" and "Left" inputs, we can progressively configure the MZI to route all the input power to the "Right" output by minimizing the power out of the "Bottom" output:

first adjust the $\Delta\phi$ phase shift to minimize the "Bottom" output power
this sets the portions of the "Top" and "Left" inputs to be in antiphase at the "Bottom" output
then adjust the phase shift to minimize the "Bottom" power again
this makes the "Top" and "Left" input portions at this output have equal and opposite amplitudes that cancel completely at the "Bottom" or "drop port" output to give zero output.

If we self-configure this way, then we are setting

$$\Delta\phi = \beta - \alpha \tag{4}$$

and

$$\Delta\theta = 2\tan^{-1}\left(|a/b|\right) \tag{5}$$

When we do this, we are setting the vector $[u_{1,1} \quad u_{2,1}]$ corresponding to the first row in the matrix $\mathsf{M}_T$ to be the Hermitian adjoint of the input vector, i.e., $u_{1,1} = a^*$, $u_{1,2} = b^*$. Hence, the self-configuration operation with some input vector $|e_{in}\rangle$ sets the first row of the matrix to be its Hermitian adjoint $\langle e_{in}|$. Equivalently, if we want to configure a mesh to represent a matrix whose first row is $\langle e_1| = [u_{1,1} \quad u_{2,1}]$, then we should self-configure it with the input vector $|e_{in}\rangle = |e_1\rangle$. This concept of being able to configure any desired matrix using input vectors that are Hermitian adjoints of the rows can be regarded as a general property of self-configuring approaches. Note too that the mathematical design of this mesh to represent a given matrix just requires simple arithmetic calculations, which is also an attribute more generally of meshes that can be set up by self-configuration.

### *A simple example – a single layer mesh of identical blocks*

We can consider the simplest possible case of this kind of architecture, which is a single layer of blocks, all set identically, as shown in Fig. 2. Though this might seem almost trivial, it could make an image sensing system that also measures angle of arrival at each pixel.

Formally, we are interested in using these blocks to detect the relative phase of the beams arriving at the two inputs to the block, with the field at the "Top" input leading that at the input labelled "Left" by some amount $\phi_a$.

For this example, we set $\Delta\phi = \pi/2$ and $\Delta\theta = \pi/2$. The choice of $\Delta\theta = \pi/2$ means that $\cos(\Delta\theta/2) = \sin(\Delta\theta/2) = 1/\sqrt{2}$, so the MZI is functioning as a 50:50 beamsplitter, and $\exp(i\Delta\phi/2) = (1/\sqrt{2})[1+i]$. The resulting mesh matrix, from Eq. (1), is

$$\mathrm{M}_T(\Delta\phi = \pi/2, \Delta\theta = \pi/2) = \frac{1}{2}\begin{bmatrix} 1+i & 1-i \\ 1+i & -1+i \end{bmatrix} \tag{6}$$

Then, presuming equal powers arriving at the two inputs, the corresponding (normalized) input vector can be written as

$$|e_{in}\rangle = \frac{1}{\sqrt{2}}\begin{bmatrix} \exp(-i\phi_a/2) \\ \exp(i\phi_a/2) \end{bmatrix} \tag{7}$$

where we have split the phase delay symmetrically between the two inputs to simplify the algebra.

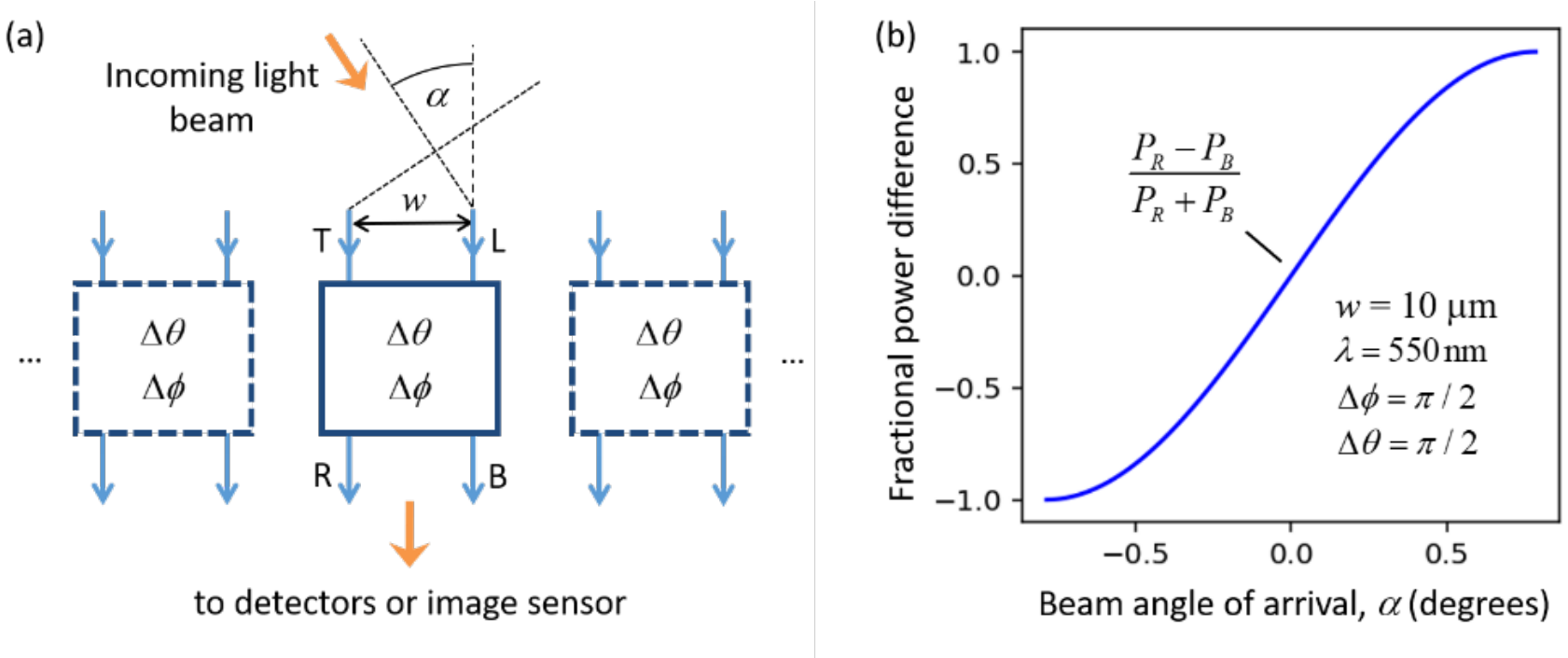

Fig. 2. (a) A single-layer mesh with blocks with identical $\Delta\phi$ and $\Delta\theta$ settings, illuminated by a beam at an angle a to the vertical. The inputs of a given block are separated by a distance w. Labels T, L, R, and B refer to the "Top", "Left", "Right", and "Bottom" ports as in Fig. 1(a). The output powers landing on detectors or image sensor pixels would allow the detection angle of arrival a of the beam. (b) The fractional difference in the powers $P_R$ and $P_B$ emerging from the R and B block outputs as a function of the arrival angle $\alpha$ of the beam.

With this matrix $\mathrm{M}_T$ as in Eq. (6) and this input vector $|e_{in}\rangle$, and writing the output powers as $P_{right} = |c|^2$ and $P_{bottom} = |d|^2$ in appropriate units, the fractional difference in the powers becomes, after some algebra

$$F \equiv \frac{P_{right} - P_{bottom}}{P_{right} + P_{bottom}} = \sin(\phi_a) \tag{8}$$

So, the phase angle between the two (equal power) inputs to the block can be simply deduced from the fractional difference $F$ in powers at the block output, giving, from Eq. (8)

$$\phi_a = \sin^{-1} F \tag{9}$$

Suppose that, for a beam whose intensity is relatively constant over the separation $w$ between the block inputs, we record the powers at the two outputs, e.g., on adjacent pixels on an image sensor. Averaging over the two outputs of the block gives a conventional image, while the difference gives us phase gradient information that we could calculate in real time or simply save for future interpretation.

Specifically, suppose that the input to this block is a beam at an angle $\alpha$ to the vertical, as shown in Fig. 2, such that the phase of the wave arriving at the "Top" input leads that arriving at the "Left" input by an amount

$$\phi_a = kw \sin \alpha \tag{10}$$

where $k = 2\pi / \lambda$ with $\lambda$ as the free-space wavelength. Then, for a monochromatic beam of wavelength $\lambda$, we can use these blocks to measure the input angles of the beam at each such block. Explicitly, from Eqs. (9) and (10), we would deduce

$$\alpha = \frac{1}{kw} \sin^{-1}\left(\sin^{-1} F\right) \tag{11}$$

For example, if we choose a separation between inputs to the block of $w = 10$ microns and a wavelength of 550 nm, then the relation between beam angle and the fractional power difference is as shown in Fig. 2(b). Smaller separation $w$ or a longer wavelength will give proportionally larger angular ranges.

Alternatively, at a fixed angle $\alpha$, we could deduce the wavelength $\lambda$ or frequency of locally monochromatic light from $F$ in a similar manner. See Supplemental document section S3.

### *Embedded binary tree*

Before proceeding to more general examples, we note that it can be straightforward to embed existing powerful and self-configuring architectures in the trapezoidal mesh. Specifically, the binary tree architecture [13,14] can be implemented just by choosing some MZI blocks to be fixed as illustrated in Fig. 3, where Layer 2 of the mesh is set in "cross" and "bar" states to perform the necessary routing to embed this version of a binary tree.

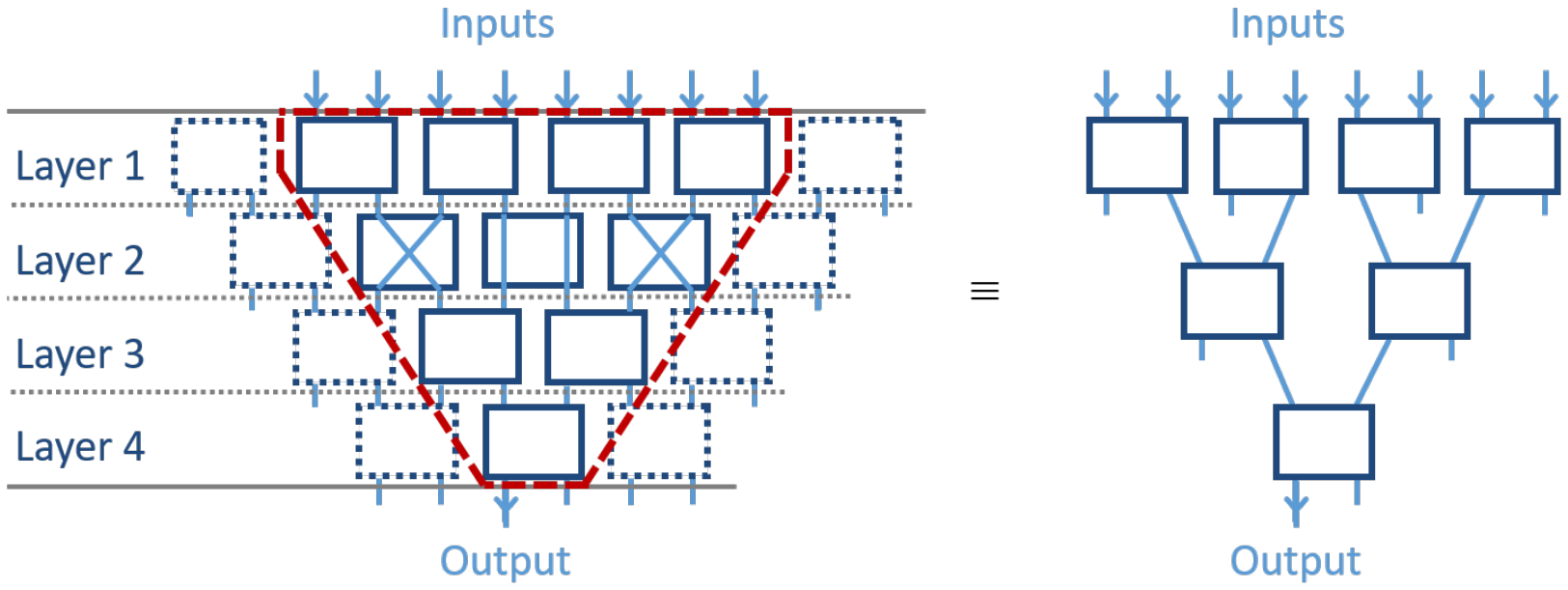

Fig. 3. Example of embedding an 8-input binary tree in a 4-layer trapezoidal mesh, using Layer 2 purely for routing in "cross" and "bar" configurations.

This 8-input binary tree architecture allows the power in any set of 8 relative amplitudes and phases as the input vector to be combined to the one chosen Output; with the Layer 2 blocks fixed as shown the necessary settings for the remaining blocks can be self-configured just by shining in the desired input vector. Running backwards by shining light backwards into the Output port allows any 8-element vector of relative amplitudes and phases to be generated [13,14]. Hence, this serves as an example of how this trapezoidal architecture is capable of some quite arbitrary linear operations. To understand the full range of operations can be performed by this trapezoidal architecture class, we proceed to a more formal analysis.

## 4. Analysis of multilayer meshes

*Constructing the full mesh matrix from block settings*

As we analyze the mesh we need the matrix that defines the linear transformation from the inputs to the outputs for the full mesh, based on the matrices for the individual $2\times 2$ blocks. For a mesh with $N$ waveguides or inputs, the effect of an individual $2\times 2$ block $p$ between adjacent waveguides or inputs $j$ and $j+1$ can be written as the $N\times N$ matrix

$$\mathsf{M}_p = \begin{array}{c} \text{outputs} \\ \\ j \\ j+1 \\ \\ \\ \end{array} \overset{\begin{array}{c}\text{inputs}\\ j \quad j+1\end{array}}{\begin{bmatrix} 1 & 0 & \cdots & & \cdots & 0 \\ 0 & \ddots & & & & \vdots \\ \vdots & & \boxed{\mathsf{M}_{Tp}} & & & \\ & & & & & \\ \vdots & & & & \ddots & 0 \\ 0 & \cdots & & & 0 & 1 \end{bmatrix}} \tag{12}$$

where $\mathsf{M}_{Tp}$ is the $2\times 2$ matrix for the block as in Eq. (1). The matrix for the entire mesh is simply the product of the matrices for all the blocks, taken in order starting from the inputs and connected according to the mesh connections. For blocks within a given layer in the mesh as in Fig. 1(c), the matrices for individual blocks in the layer can be multiplied in any order because they do not interact with one another [26]; as a result, these can also be written as a set of blocks in the diagonal of one matrix $\mathsf{M}_{Lr}$ for layer $r$. Then we can multiply those layer matrices $\mathsf{M}_{L1}$, $\mathsf{M}_{L2}$, … to get the overall matrix $\mathsf{U}_M$ for the mesh [26] – e.g., for the four-layer mesh of Fig. 1, we would have

$$\mathsf{U}_M = \mathsf{M}_{L4}\mathsf{M}_{L3}\mathsf{M}_{L2}\mathsf{M}_{L1} \tag{13}$$

*Setting up the mesh*

We see that, if we set the mesh up by self-configuring each block to give zero at one of its outputs, we can either physically set up a programmable mesh directly by shining in vectors of interest or equivalently calculate the matrix for the entire mesh in this way using the resulting expressions Eqs. (4) and (5) and the larger matrix construction as in Eqs. (12) and (13).

For this analysis, we label the inputs, outputs, blocks and configuration vectors as in Fig. 4. We discuss this mesh as if we were self-configuring it with vectors that correspond to the various rows of the matrix it is to implement. If we can devise such a self-configuring algorithm, then it is also straightforward to calculate the required settings of the blocks because those emerge as we perform the self-configuration (physically or mathematically).

Incidentally, not all mesh architectures appear to support self-configuration. Meshes formed from one of more cascaded so-called "self-configuring layers", such as diagonal lines or binary trees [13–15,19,20], do support self-configuration just using the vectors of interest – that is, the ones that are to be routed entirely to the output of that layer and that correspond to the Hermitian adjoints of the rows of the matrix to be implemented. Rectangular meshes [27], though they can be set up using a larger set of input vectors [25], do not appear to support self-configuration. Self-configuring layers have the topological property that there is one and only one path through the mesh layer from its output to each input [13], but a rectangular mesh has many such paths. We can see that the trapezoidal architecture also does not have this topological feature. For example, there are multiple paths from Output 1 back to Input 3, passing through blocks B32 or B31 in Layer 3, or blocks B21 or B22 on Layer 2. Hence, we might presume this network cannot be self-configured. However, as we show below, the network can still be set up using only single-parameter power maximizations or minimizations and using only the vectors that correspond to the rows of the matrix to be configured, though we may have to use those vectors more than once in the algorithm. So, in this broader sense, the network can be self-configuring. Also, once the main trapezoid (as shown in Fig. 1) is set up, we will see that each additional diagonal line on the right (e.g., the one starting with block B18) is set up just by self-configuring that line with the one additional input Vector for that diagonal line.

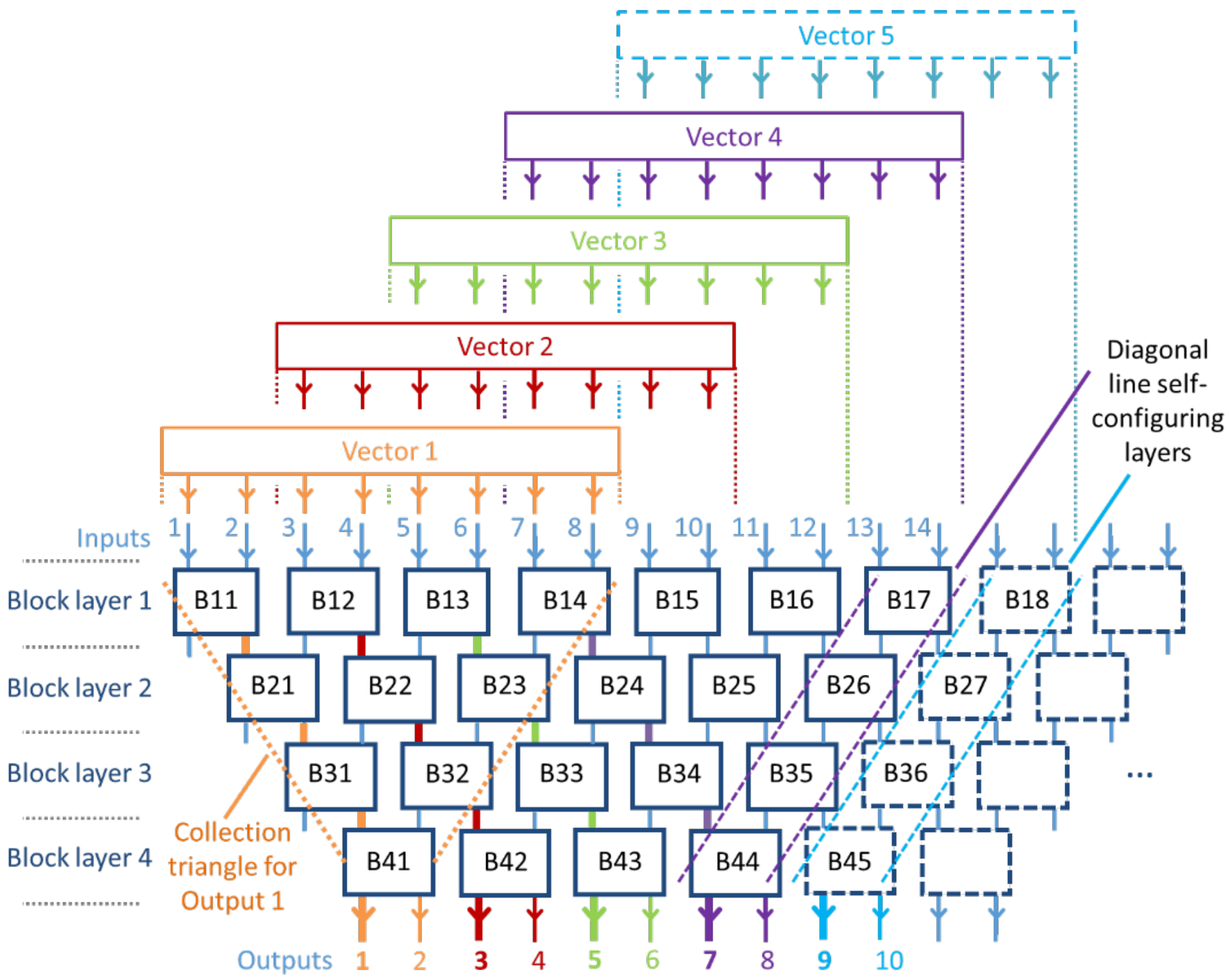

Fig. 4. Example set of 4 layers of $2\times2$ blocks – B11, B12, etc. – that can be programmed by shining in 8-element vectors of amplitudes – Vector 1, Vector 2, etc. – just by successive single-parameter power maximizations or minimizations. Appropriately orthogonal Vectors 1 through 4 are sufficient to set the blocks to route Vectors 1 through 4 one by one each to its corresponding Output 1, 3, 5, or 7. Because of the mesh construction, each Output can only collect light arriving within its collection triangle, as shown for Output 1. The system can be extended laterally and progressively by subsequent vectors, such as Vector 5, each orthogonal to the preceding 3 vectors. Those subsequent Vectors are each sufficient, by self-configuration of the corresponding diagonal line, similarly to route all their power to the corresponding outputs – e.g., Output 9 for Vector 5.

In describing the algorithm for setting up (or calculating the settings of) the mesh, we presume or pretend that we have (mostly transparent) photodetectors at the outputs of each block so we can monitor the powers at the outputs of each block [14,15]. However, even if we only have detectors on the overall output ports, we can effectively use those for this purpose if we can temporarily set the various other blocks as needed in their "cross" state (i.e., inputs swapped to the outputs) or "bar" state (i.e., inputs straight through to the outputs) to route the block outputs temporarily as needed to detectors at the outputs of the mesh [13–15]. Finding "cross" and "bar" states of blocks within the mesh is straightforward if we can illuminate the individual mesh inputs; then simple maximization or minimization of the detected output power can readily and progressively find the required settings for "cross" or "bar" states of all the blocks.

In the example 4-layer mesh in Fig. 4, we see immediately that only the 14 numbered inputs can contribute to light at the first 8 output ports; light at other input ports has no path by which it can get to any of these 8 output ports. Also, only the 8 ports in the collection triangle for a given output can contribute to that output. So immediately we can write down the form of the matrix that maps the amplitudes at those 14 input ports to those 8 output ports, which is therefore

$$\mathsf{U}_M = \begin{bmatrix} u_{1,1} & u_{1,2} & u_{1,3} & u_{1,4} & u_{1,5} & u_{1,6} & u_{1,7} & u_{1,8} & 0 & 0 & 0 & 0 & 0 & 0 \\ u_{2,1} & u_{2,2} & u_{2,3} & u_{2,4} & u_{2,5} & u_{2,6} & u_{2,7} & u_{2,8} & 0 & 0 & 0 & 0 & 0 & 0 \\ 0 & 0 & u_{3,3} & u_{3,4} & u_{3,5} & u_{3,6} & u_{3,7} & u_{3,8} & u_{3,9} & u_{3,10} & 0 & 0 & 0 & 0 \\ 0 & 0 & u_{4,3} & u_{4,4} & u_{4,5} & u_{4,6} & u_{4,7} & u_{4,8} & u_{4,9} & u_{4,10} & 0 & 0 & 0 & 0 \\ 0 & 0 & 0 & 0 & u_{5,5} & u_{5,6} & u_{5,7} & u_{5,8} & u_{5,9} & u_{5,10} & u_{5,11} & u_{5,12} & 0 & 0 \\ 0 & 0 & 0 & 0 & u_{6,5} & u_{6,6} & u_{6,7} & u_{6,8} & u_{6,9} & u_{6,10} & u_{6,11} & u_{6,12} & 0 & 0 \\ 0 & 0 & 0 & 0 & 0 & 0 & u_{7,7} & u_{7,8} & u_{7,9} & u_{7,10} & u_{7,11} & u_{7,12} & u_{7,13} & u_{7,14} \\ 0 & 0 & 0 & 0 & 0 & 0 & u_{8,7} & u_{8,8} & u_{8,9} & u_{8,10} & u_{8,11} & u_{8,12} & u_{8,13} & u_{8,14} \end{bmatrix} \tag{14}$$

Note that this matrix $U_M$ has 14 columns, each of which corresponds to one of the inputs, 1 through 14, to the first layer of blocks, and 8 rows, each of which corresponds to one of the outputs 1 through 8 of the last layer of blocks. Each row has only 8 non-zero elements, which corresponds to the fact that only those 8 inputs contribute, in the “collection triangle” of blocks in Fig. 4, to the corresponding output. We can also conveniently choose the various rows to be normalized; writing a row in Dirac notation as $\langle \psi_n |$ for the *n*th row of (complex) matrix elements, then normalization corresponds to $\langle \psi_n | \psi_n \rangle = 1$ (where $| \psi_n \rangle$ is the mathematical column vector that is the Hermitian adjoint (conjugate transpose) of $\langle \psi_n |$). In what follows, we show how we can set up the blocks to implement any such matrix (with orthogonal and normalized rows). For simplicity of discussion, we refer to $\mathsf{U}_M$ as being unitary; as a matrix it technically is not (it has different numbers of rows and columns) and can be called a semi-unitary or partially unitary matrix, though as an operator it can be rigorously unitary (See Supplemental document Section S4).

We can immediately deduce some properties of this matrix. From the physics of the system, we see that the matrix must preserve power; the network between inputs and outputs is nominally lossless. As a consequence, the matrix row vectors are orthogonal to one another (i.e., $\langle \psi_n | \psi_m \rangle = \delta_{nm}$) (15)

This is because the corresponding outputs, being physically separated, are necessarily physically orthogonal, which means the input vectors corresponding to them must be orthogonal – lossless physical networks must preserve orthogonality, otherwise we could violate the second law of thermodynamics (see the Supplemental document Section S1 of Ref. [9]).

Because of this orthogonality of the rows and their normalization, quite generally a (unit power) input vector of amplitudes $| \psi_n \rangle$ would be combined to give unit output at Output *n*. So, as mentioned above, for an input vector of amplitudes $| \psi_n \rangle$ to give an output only in Output *n*, the corresponding matrix row is $\langle \psi_n |$ (16)

We note also that, once we have set up the blocks to route to a specific output, such as the blocks in the collection triangle for Output 1 in Fig. 4, and hence having set the first row $\langle\psi_1|$ of the matrix $\mathsf{U}_M$ , we have also defined exactly what the second row $\langle\psi_2|$ of the matrix must be (at least within an overall phase factor). So, choosing the *n*th matrix row $\langle\psi_n|$, where Output *n* corresponds to a "left" output from a $2\times2$ block, determines the $(n+1)$ row $\langle\psi_{n+1}|$ by simple calculation, where Output $n+1$ is the "right" output of the same block

(17)

We will refer to this predetermined vector $\langle\psi_{n+1}|$ as the "complement" of the corresponding vector $\langle\psi_n|$ from which it can be calculated. Explicitly, if we calculate the mesh matrix $\mathsf{U}_M$ , even if only for the "collection triangle" of blocks, then this complement vector $\langle\psi_{n+1}|$ is just the next ( $n+1$ th) row of the matrix, and the input vector $|\psi_{n+1}\rangle$ that would route to the $n+1$ th output is just the Hermitian adjoint $|\psi_{n+1}\rangle$ of the $n+1$ th row ( $\langle\psi_n|$ ) of $\mathsf{U}_M$ .

To see why this complementary vector $\langle\psi_{n+1}|$ is necessarily predetermined once we have set the mesh to route $|\psi_n\rangle$ to output *n*, consider the case of $n=1$ , which has all the power in input Vector 1, $|\psi_1\rangle$, routed to output 1. Now we shine light backwards into output 2; the resulting vector of amplitudes $|\gamma\rangle$ that emerges backwards from the inputs is completely defined by the settings of the blocks within the collection triangle for Output 1, and those settings have already been chosen as we routed all of Vector 1 ( $|\psi_1\rangle$ ) to Output 1. The phase conjugate of vector $|\gamma\rangle$ would be the input vector that would all be routed to Output 2 [13,25], which is, of course the vector $|\psi_2\rangle$, which is just the Hermitian adjoint of the second matrix row, $\langle\psi_2|$. So, that second matrix row is set automatically once we configure the mesh to route $|\psi_1\rangle$ to be routed all to Output 1. We could refer to these corresponding complementary input vectors as Vector $1'\equiv|\psi_2\rangle$, Vector $2'\equiv|\psi_4\rangle$, and so on.

This continues for row 4 of the matrix being implicitly determined once we have set row 3 of the matrix, and so on. So only every second row (that is, rows 1, 3, 5, etc.) is set independently (other than possibly an overall phase). This behavior is consistent with the rows being orthogonal; for example, matrix rows 1 and 2 must be orthogonal anyway since they would correspond to input vectors that route to different output ports.

If we add diagonal lines of blocks (as in blocks 18, 27, 36, and 45 in Fig. 4), then we just add additional pairs of rows to the bottom of this matrix (so adding 2 rows to the matrix), shifting the non-zero elements in each pair to the right by two columns (and so also increasing the number of columns by 2).

### *Algorithm for setting the blocks by self-configuration*

Now we examine the algorithm that will allow us to configure the entire mesh based only on the input vectors corresponding to these rows 1, 3, 5, etc. of the desired mesh matrix; we do this by example for a 4 layer mesh, with the generalization for different numbers of layers then being straightforward.

Given that we want the mesh of blocks to implement some specific matrix $\mathsf{U}_M$, we can consider the input vectors of amplitudes $|\psi_1\rangle$ (Vector 1), $|\psi_3\rangle$ (Vector 2), $|\psi_5\rangle$ (Vector 3), and $|\psi_7\rangle$ (Vector 4) that correspond to the desired matrix rows $\langle\psi_1|$, $\langle\psi_3|$, $\langle\psi_3|$ and $\langle\psi_7|$. Physically, these will be the input vectors whose powers are to be routed respectively to Outputs 1, 3, 5, and 7. Now we construct an algorithm for setting the blocks to achieve this, using just these input Vectors and the self-configuring algorithm for the relevant $2\times2$ blocks.

Note first that there is a total of 10 blocks that need to be set in routing Vector 1 to Output 1 – specifically B11 to B14, B21 to B23, B31, B32, and B41. However, the Vector 1, which has amplitudes defined by 8 complex numbers, each of which is defined by 2 real numbers, does not have enough degrees of freedom to set the 20 real parameters required for 10 $2\times2$ blocks if these are to have their most general possible setting. As a result, the setting of these blocks is requires the use of more than just the amplitudes of Vector 1. In the algorithm we construct, we will also use Vectors 2, 3 and 4. Though we may have to use these Vectors more than once in the configuration, there is a progressive sequence of using these Vectors and setting specific block outputs to zero that can be used to configure the mesh completely. In this broader sense, this architecture can be self-configuring – we can set it up just using the vectors we want to be routed to specific outputs, though we may have to use a given vector more than once in the algorithm, at least in configuring this main trapezoid of blocks.

Note as in Fig. 4 that successive Vectors shine into different sets of inputs, translated successively to the right by two waveguides (or one block). Note too that these vectors must all be mathematically orthogonal to one another; otherwise they could not be routed by the mesh to separate (and hence orthogonal) outputs.

We state the algorithm based on power minimization in an output port of a block; it could also be run by power maximization in the other output port of the block. The basic trick in this algorithm is that we exploit the situations where we know the input fields for a given block must be routed to just one output of that block. The configuration of the blocks proceeds by a succession of such steps. We can view this as a physical process that configures the blocks, but we can also use it mathematically as a design process.

Because of the finite collection triangle for a given output, we can deduce immediately that, when shining in Vector 1, which is the vector of input amplitudes that is to be routed to Output 1, B11 must route all its input power to its "right" output, and similarly block B14 must route all its input power to its "left" output. (Note: here we use the usual meaning of "right" and "left", though these correspond to "Bottom" and "Right" respectively in the formal notation of a $2\times2$ block in Fig. 1(a).) These kinds of operations will form the basis for the construction of the algorithm. Having done that, we can then use Vector 2 to set B12 to route all its power to its right output, and Vector 3 to set B13 to route all its power to its right output. Returning to Vector 1, we can now set block B21 to route all its power to its right output. We can continue in a similar fashion to set all the blocks in the "collection triangle" for block B41, hence routing all of Vector 1 to output 1. Proceeding in this way just with these four inputs Vectors, we can route all the powers in each of Vectors 1 to 4 to their corresponding single outputs 1, 3, 5, and 7, thus setting all 10 blocks (B11, B12, B13, B14, B21, B22, B23, B31, B32, and B41) in the trapezoid of Fig. 1. A detailed algorithm is given in Supplemental document section S5.

Now that we have configured this trapezoid of blocks, it is straightforward to extend this arbitrarily to the right by adding diagonal lines of blocks as in Vector 5 in Fig. 4. We can configure the corresponding diagonal line of blocks B18, B27, B36, B45 now just by progressively zeroing out the power in the right output port of each such block in sequence with Vector 5 as the input. Vector 5 must, of course, be orthogonal to Vectors 2, 3, and 4 (and it will automatically be orthogonal to Vector 1 because it does not overlap with it). We can proceed similarly for any further diagonal lines added on the right.

An alternative configuration algorithm, based only on self-configuring diagonal lines, is given in Supplemental document Section S6. In this approach, we discard the first several inputs and outputs of the mesh but get a simpler algorithm as a result.

Hence, we can self-configure a layered structure like this of arbitrary width, just using input vectors that are the Hermitian adjoints of the desired (odd-numbered) matrix rows. This gives us a general and simple design method to represent any banded matrix of the form of $\mathsf{U}_M$ (with orthogonal rows) or its extensions with additional columns and rows by adding diagonal lines of blocks. As mentioned, we can also use this algorithm mathematically to calculate the block settings in our design even if we never perform physical power minimizations or maximizations.

## 5. Example multilayer mesh – implementing a filter based on Daubechies wavelet coefficients

A good example to show the larger capabilities of the trapezoidal architecture is design one to implement a filter (here in the spatial domain) based on the coefficients associated with a Daubechies wavelet. Such wavelets are widely used in signal processing. Specifically, we can consider the db4 wavelet, which can correspond to an 8 tap finite-impulse-response filter. The decomposition wavelet coefficients for that filter (obtained, for example, through the PyWavelets [28] pywt.Wavelet('db4').dec_hi function) are, approximately

db4wv = [-0.2304, 0.7148, -0.6309, -0.028, 0.187, 0.0308, -0.0329, -0.0106] (18)

with the corresponding decomposition scaling coefficients (through the PyWavelets pywt.Wavelet('db4').dec_lo) function

db4sc = [-0.0106, 0.0329, 0.0308, -0.187, -0.028 0.6309, 0.7148 0.2304] (19)

(Note, incidentally, that the scaling coefficients are the wavelet coefficients in reverse order, with alternate signs switched.) Using the wavelet coefficients as the taps in such a filter leads to a “high pass” filter, and the scaling coefficients lead to a “low pass” filter. We can configure a four-layer mesh to implement a one-dimensional spatial filter with the db4wv as the configuration vector.

An important property of the vector db4wv (and also of db4sc) is that it is orthogonal to versions of itself shifted by any multiple of 2 inputs. Writing these vectors in an extended form (as will be appropriate as rows that are 14 “input” elements long to correspond to the trapezoidal architecture for 4 layers), the vector

db4scx(0) = [-0.2304, 0.7148, -0.6309, -0.028, 0.187, 0.0308, -0.0329, -0.0106, 0,0,0,0,0,0]

is orthogonal to a version of itself shifted to the right by two positions, i.e., the vector

db4scx(2) = [0, 0, -0.2304, 0.7148, -0.6309, -0.028, 0.187, 0.0308, -0.0329, -0.0106, 0,0,0,0]

and to the vectors db4scx(4) and db4scx(6) that are similarly shifted by 4 and 6 positions, respectively. In fact, all 4 of these vectors are orthogonal to one another. (If we were to extend these vectors by adding more zeros to correspond to the inputs to a wider trapezoid with added diagonal lines, any further such shifts by two positions would also be orthogonal because the non-zero elements would not overlap with each other.)

Hence, because of these orthogonalities, these vectors db4scx(0), db4scx(4), db4scx(4) and db4scx(6) could be used as Vectors 1 to 4 to configure the trapezoidal mesh, as in Fig. 4 above, and further such shifted versions could continue to configure a further extended trapezoidal mesh with added diagonal lines.

Because these configuration vectors are simply shifted versions of one another, as a result, all the blocks in a given layer are configured identically. We can see this if we think of the collection triangles for each of the outputs 1, 3, 5, etc. Since the vectors that configure for each of these shifted outputs are identical, so also must the settings of each such collection triangle be identical. The only way this can be the case is if all the blocks in a given layer are identical. We can prove this more mechanistically by following through the configuration algorithm. For example, the first two elements Vector 1 must lead to the output of block B11 being routed to its right output; but the first two elements of Vector 2 (which is just a shifted version of Vector 1) must similarly lead to the output of block B12 being routed to its right output, so therefore blocks B11 and B12 are set identically, and so on. The resulting mesh design to implement this db4 (spatial) filter is shown in Fig. 5.

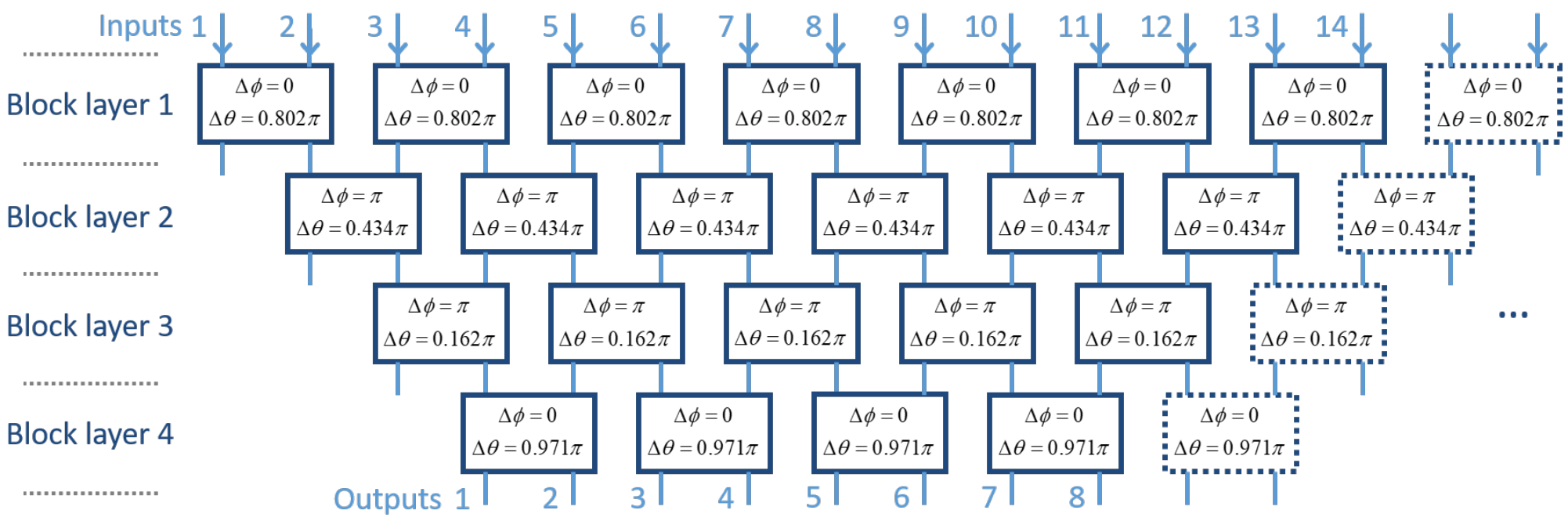

Fig. 5. Settings of blocks to implement filters based on db4 wavelet coefficients, with all blocks in layer 1 set to $\Delta\phi = 0$, $\Delta\theta \simeq 0.802\pi$, all blocks in layer 2 set to $\Delta\phi = \pi$, $\Delta\theta \simeq 0.434\pi$, all blocks in layer 3 set to $\Delta\phi = \pi$ $\Delta\theta \simeq 162\pi$, and all blocks in layer 4 set to $\Delta\phi = 0$, $\Delta\theta \simeq 0.971\pi$.

One intriguing point is that not only does this mesh perform a db4 wavelet-based filter to outputs 1, 3, 5, 7, etc., it also performs the corresponding db4 scaling-based filter to outputs 2, 4, 6, 8, etc. That is, the complementary vector to db4wv is actually db4sc.

The mesh matrix for the trapezoid is, explicitly (rounding to 2 figures for compactness),

$$\mathsf{U}_M = \begin{bmatrix} -.23 & .71 & -.63 & -.03 & .19 & .03 & -.03 & -.01 & 0 & 0 & 0 & 0 & 0 & 0 \\ -.01 & .03 & .03 & -.19 & -.03 & .63 & .71 & .23 & 0 & 0 & 0 & 0 & 0 & 0 \\ 0 & 0 & -.23 & .71 & -.63 & -.03 & .19 & .03 & -.03 & -.01 & 0 & 0 & 0 & 0 \\ 0 & 0 & -.01 & .03 & .03 & -.19 & -.03 & .63 & .71 & .23 & 0 & 0 & 0 & 0 \\ 0 & 0 & 0 & 0 & -.23 & .71 & -.63 & -.03 & .19 & .03 & -.03 & -.01 & 0 & 0 \\ 0 & 0 & 0 & 0 & -.01 & .03 & .03 & -.19 & -.03 & .63 & .71 & .23 & 0 & 0 \\ 0 & 0 & 0 & 0 & 0 & 0 & -.23 & .71 & -.63 & -.03 & .19 & .03 & -.03 & -.01 \\ 0 & 0 & 0 & 0 & 0 & 0 & -.01 & .03 & .03 & -.19 & -.03 & .63 & .71 & .23 \end{bmatrix} \tag{20}$$

So, if we were to present some signal, such as a row of a set of image pixels, to the inputs to the mesh, the odd outputs would be that signal filtered by the db4wv kernel and the even outputs by the db4sc kernel.

If we made a wide version of this mesh by adding many diagonal lines, or equivalently had some wide stacked metastructure to implement this function, then when shining a line of inputs, such as a line of pixels from an image, onto the mesh inputs, the mesh effectively would perform the convolution of that line of inputs with these db4 coefficient kernels, with the results for db4wv and db4sc kernel appearing at alternating outputs of the mesh.

The example of the mesh to implement these db4-based filters is obviously periodic when extended laterally. It repeats as a set of diagonal lines that could be viewed as some kind of unit cell of a crystalline structure, and we could refer to this as a simple periodic mesh (with all diagonal lines the same). It is possible also to construct more complex periodic structures, for example with groups of 4 diagonal lines repeating in a larger parallelogram unit cell. We will not pursue such periodic structures further here, but the formalism for analyzing them as a kind of "forward" photonic crystal is given in the Supplemental document, section S7, with associated configuration algorithms in Supplemental document sections S8 and S9.

## 6. Approaches to manufacturable multilayer structures

As mentioned above, this architecture could be one novel way to work with large planar silicon photonic MZI meshes. Making deep meshes with the number of MZI blocks in each layer comparable to the number of layers is challenging because the MZIs are typically much longer, e.g., by a factor of ~ 10, than their width, so such a circuit would be ~10 times longer than its width. This architecture allows us to contemplate circuits with relatively smaller depth in terms of the number of layers of MZIs so that can exploit the relatively square shape of a full lithographic reticle and a large number of inputs. Such circuits could be fabricated with current technology.

More speculatively, we could contemplate separately fabricating standardized metasurfaces that could then be stacked to create the overall structure, as in Fig. 6. Such a fabrication could, of course, be challenging, but it is likely more feasible than complex metastructures with tens or more layers of subwavelength dimensions that also have to be stacked with subwavelength precision in all three dimensions [12]. While subwavelength precision in fabrication laterally within a layer is achievable because of advanced lithographic techniques, precise vertical thickness is more difficult to achieve with comparable precision. The forward-only nature of the present concept means that such vertical precision is not required because there are no reflections inside the structure between the layers that could lead to interlayer interference effects.

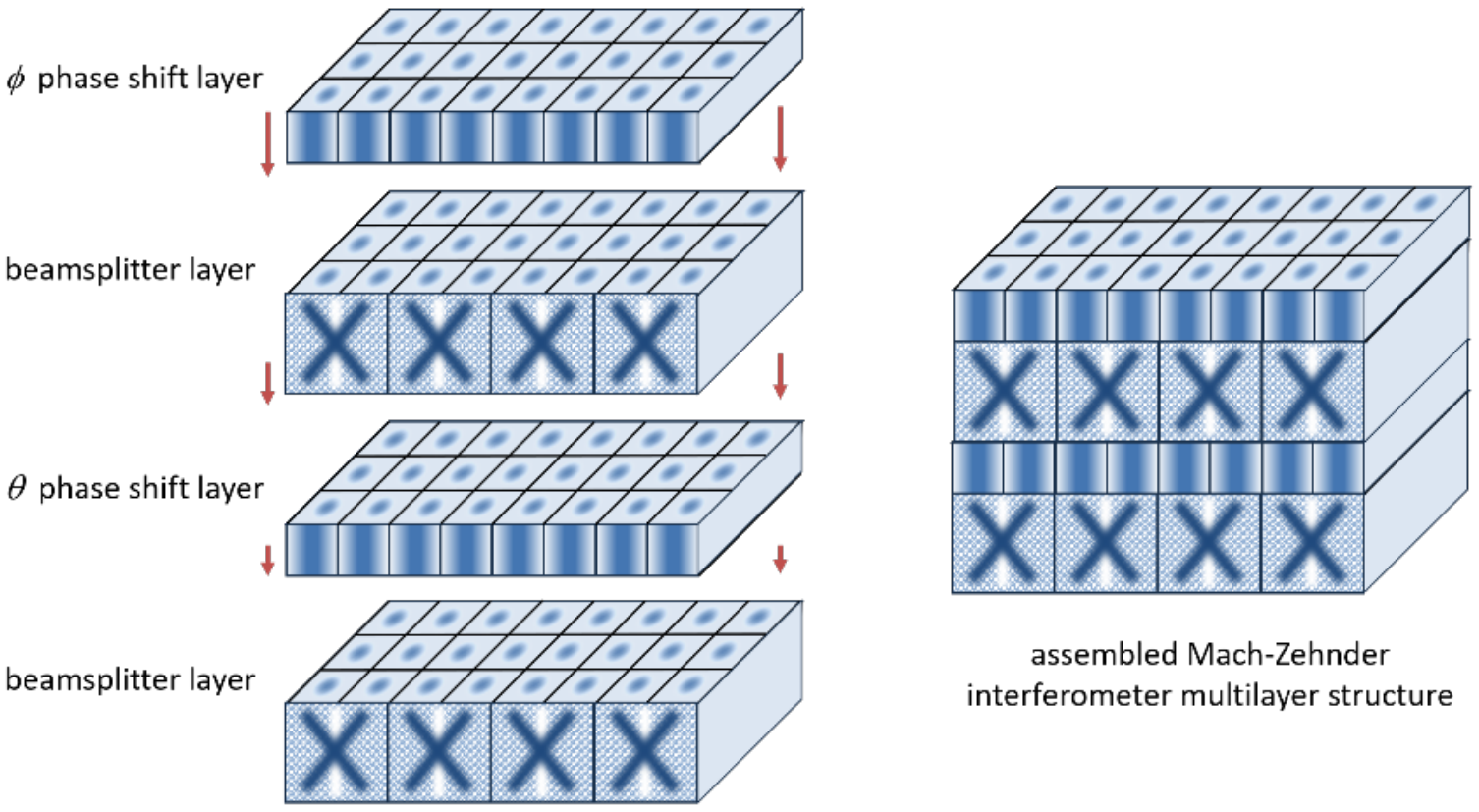

Fig. 6. Conceptual assembly of a compound layer of Mach-Zehnder interferometers by stacking phase shift layers and 50:50 beamsplitter layers.

Note too that this approach can work with standardized layers, especially for the beamsplitter layers. The design of the function of the mesh is then given entirely by the phase shift layers, which can likely be designed with some precision using metasurface techniques, even if all such phase shifts elements are to have the same overall thickness to allow a uniform layer thickness for stacking.

Other key challenges with this approach are, first, that these layers should all be designed so they have little or no reflection once stacked and, second, that the approach needs layers corresponding to $2\times2$ beamsplitters. Though many metasurface beamsplitters have been proposed [29], it appears that most and possibly all such work has targeted 1 to 2 ("Y") or 1 to $N$ splitters. Such splitters can be made with a single thin layer. However, a $2\times2$ beamsplitter necessarily has an overlapping nonlocality [9] of 2 (see Supplemental document section S1) which strongly suggests that some thickness capable of supporting two channels in a given direction is required for any such design. The design of this layer of $2\times2$ metasurface beamsplitters appears to remain an interesting challenge. Fig. 7 sketches various conceptual or actual approaches to such beamsplitters

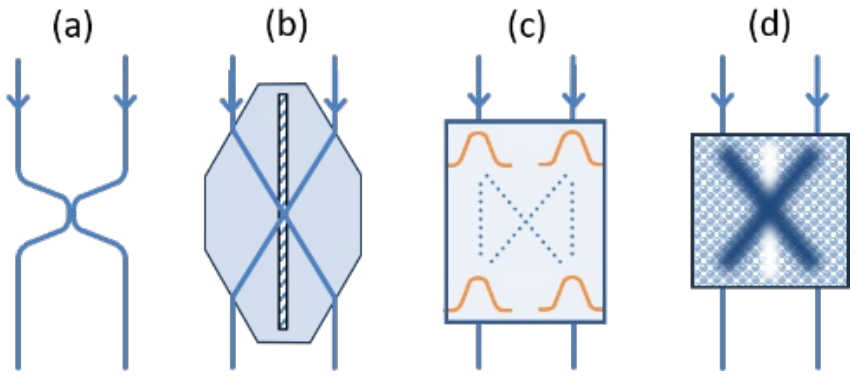

Fig. 7. Illustration of several approaches to beasmplitters for use in mesh layers. (a) A waveguide splitter operating by evanescent coupling, as in silicon photonics. (b) A hypothetical glass or dielectric beamsplitter structure with parallel inputs and outputs. (c) A multimode interference coupler. (d) A figurative notation for a beamsplitter, possibly made from some metastructure design.

In general with metasurfaces, if we stack them very closely, even with forward-only propagation, it is possible we could have evanescent coupling between layers, including coupling backwards, which would prevent the simple "forward-only" analysis and behavior here. However, this could be easily avoided just by allowing a small space or adding a thin spacer layer between successive beamsplitter or phase shifter layers in the structure.

The approaches discussed so far, other than for planar waveguide structures, are not obviously programmable after manufacture. Even if we had materials in the phase shift layers that were themselves potentially programmable, such as some electro-optic material, there is no obvious way to get the control signals into some large number of such phase shifters or other controllable elements, especially if they are in multiple different layers of the structure.

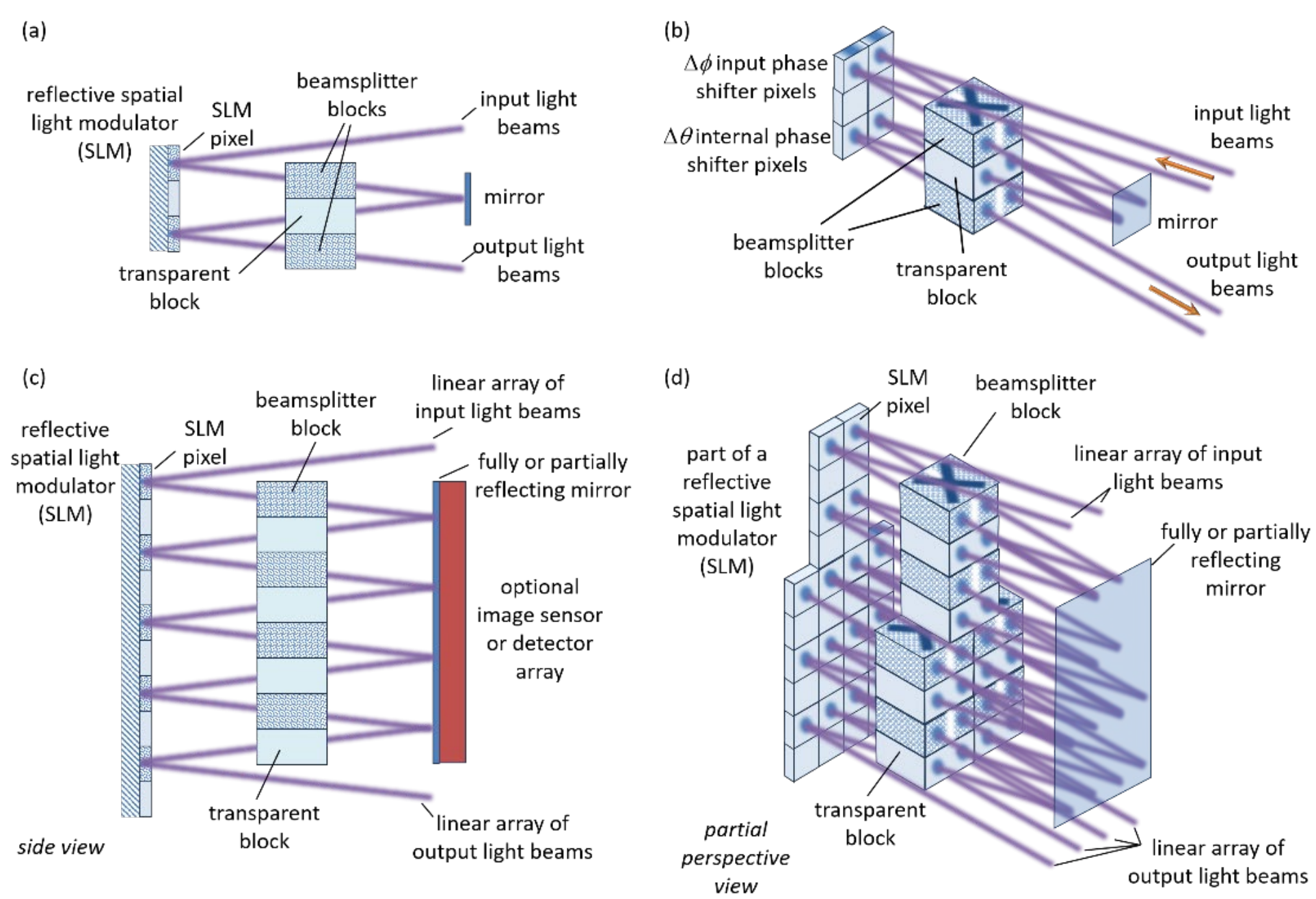

Fig. 8. Sketches of an approach to a programmable architecture, using spatial light modulator pixels as the phase shifters. (a) and (b) A stack of two beamsplitter blocks spaced by a transparent block together with two pairs of spatial light modulator (SLM) pixels form a complete MZI block. (c) and (d) A linear array of light beams, corresponding to the input light beams in Fig. 1, is processed by the stack of MZIs. Only 2 layers of MZIs are shown. The perspective view shows

only a part of the overall structure, corresponding to one MZI block (such as block B12) in the first layer (the upper column of blocks) feeding two MZI blocks (such as blocks B21 and B22) in the second layer. Only 2 input beams and 4 output beams out of the larger set are shown. The full structure would have a complete solid layer of MZI blocks and transparent blocks in 2D array.

Fig. 8 shows a concept where the architecture is folded so that it can use a spatial light modulator to provide programmable phase shifts. Fig. 8 (a) and (b) show a single MZI block. (For a fixed, non-programmable device, the spatial light modulator could also be replaced with a set of fixed phase shifts, such as thin plates of different thicknesses.) Fig. 8 (c) and (d) show a larger collection of blocks forming part of the larger architecture. If the mirror on the right is made mostly but not entirely reflecting, an image sensor or detector array on the right can monitor the powers at the ouptuts of the MZI blocks so as to provide the signals for self-configuring algorithms to set up or adapt the mesh. This approach would require just one 2D array of fixed 50:50 beamsplitters, so fabrication of the overall structure could be relatively straightforward. Optionally, appropriate lensing could be add to these blocks to keep the beams from diverging as they pass through the structure.

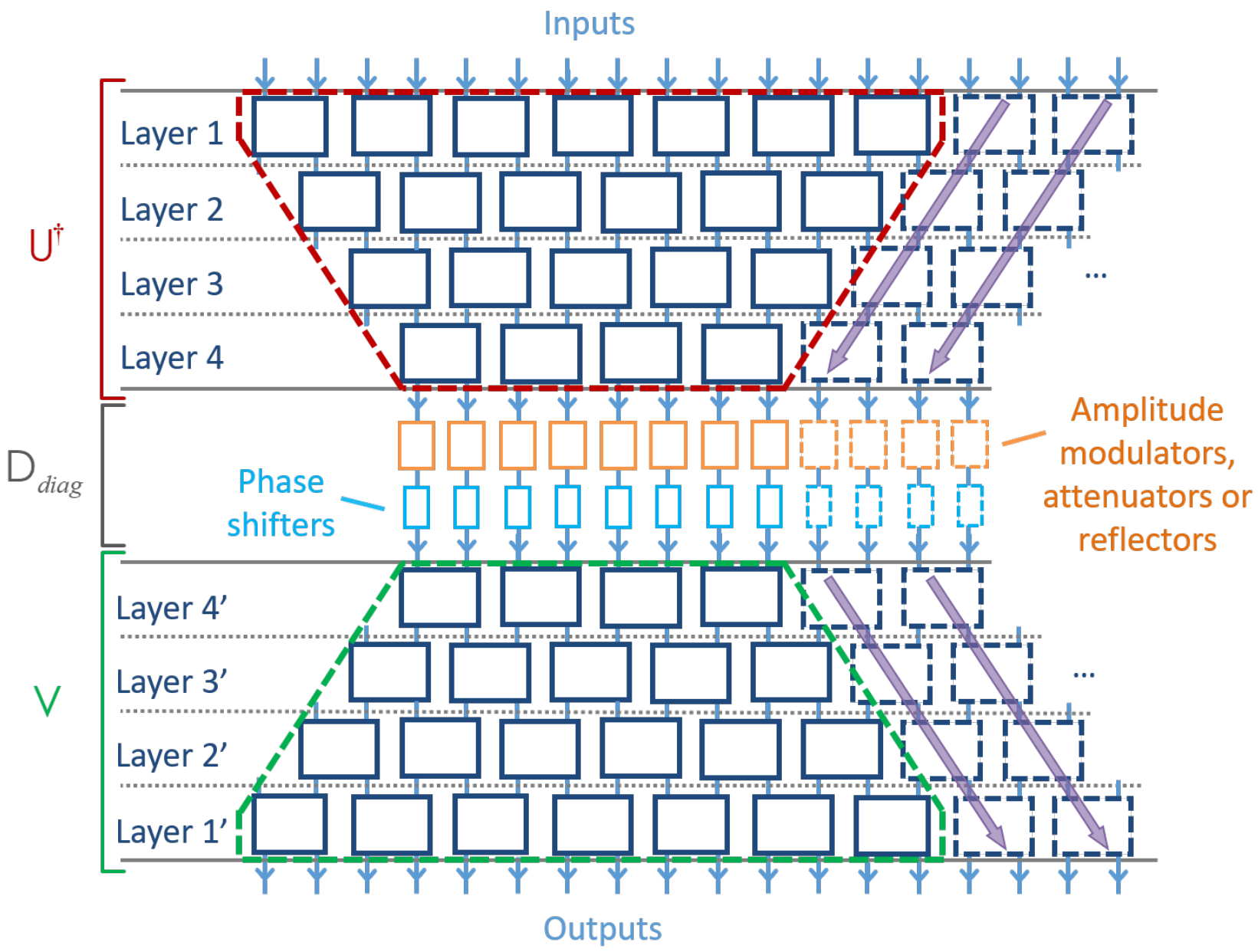

Fig. 9. Singular-value decomposition architecture for non-unitary multi-layer structures. Layers 1 to 4 give the first unitary transformation $U^{\dagger}$ . Amplitude modulators, attenuating absorbers, or reflectors, together with phase shifters give the diagonal matrix of (complex) singular values $D_{diag}$ . Layers $4'$ to $1'$ give the second unitary transformation $V$ . Together these make the non-unitary matrix through its singular value decomposition form $D = VD_{diag}U^{\dagger}$ .

## 7. Extensions

### *Non-unitary meshes*

So far, we have discussed unitary forward-only meshes. Many transformations may, however, be represented by non-unitary matrices, which will in turn require non-unitary meshes to implement them. For example, operations such as spatial differentiation are generally non-unitary. To implement non-unitary matrices, we can employ a similar strategy to that introduced in Ref. [15], which constructs the non-unitary matrix through a product $D = VD_{diag}U^{\dagger}$ , where $U$ (and hence also its Hermitian adjoint $U^\dagger$) and $V$ are unitary matrices and $D_{diag}$ is a diagonal matrix. This is constructing a non-unitary matrix through its singular-value decomposition (SVD), with the diagonal matrix elements being the (generally complex) singular values. Physically, this defines a set of orthogonal input vectors – the columns of $U$ – that map one-by-one, with connection amplitudes given by the corresponding singular value, to the corresponding column of $V$, with those columns being a corresponding set of orthogonal output vectors. In our case, the resulting matrix would be a banded diagonal non-unitary matrix.

Fig. 9 shows one way of implementing such a non-unitary forward-only layered mesh. Two layered meshes can each represent one of the unitary matrices $U^\dagger$ and $V$. The diagonal matrix $D_{diag}$ of singular values, located between the two unitary meshes, can be implemented by a set of attenuators, reflectors or active amplitude modulators to set the magnitude of the singular values, followed by phase shifts or active phase shifters to set their phase. If the singular-value coupling strengths are implemented using reflectors, the resulting "dropped" light amplitude is reflected back out of the top of the structure. As long as there are no other reflections in the structure, this will not compromise the "forward-only" behavior of the structure.

### *Spectral devices*

If we illuminate the inputs of the mesh at an angle, it becomes spectrally sensitive. We briefly discussed this for the case of one block in Supplemental document section S3S1. With multiple inputs in our architecture, illuminating with a plane wave at an angle gives a similar equal phase delay increment per input to that in an arrayed waveguide grating. Recently, spectral devices exploiting this set of delays at the input to an interferometer mesh have been proposed [22] and demonstrated [23]. This trapezoidal mesh could exploit many of the same kinds of programmable spectral functions just by illuminating it with a plane wave at an angle.

### *Two-dimensional arrays*

So far, we have only considered one-dimensional arrays or collections of them. There are many possibilities for two-dimensional arrays. We show one example in Fig. 10, appropriate for a single layer of blocks. This would allow the detection or measurement of phase gradients or beam angles in two orthogonal directions.

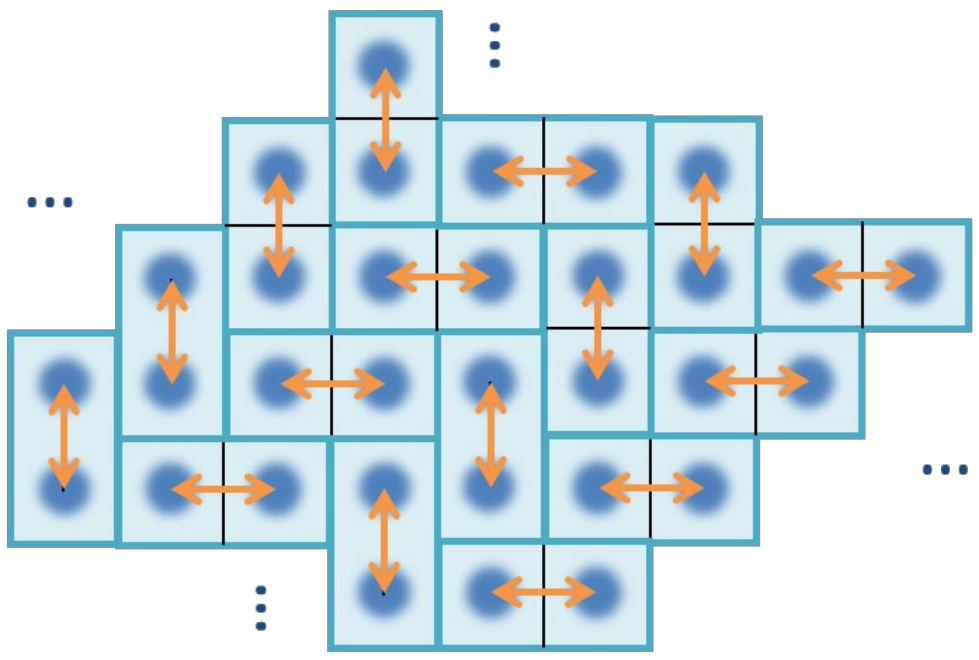

Fig. 10. Top view of the inputs to a single layer of blocks arranged in alternating "horizontal" and "vertical" orientations, as emphasized by the arrows.

## 8. Discussion

I have argued that there is a way of designing complex optical structures made from multiple successive layers, even for layers of arbitrary width or area. The capability, architecture and design complexity scale only proportionately with the number of layers in the structure. A key feature and requirement of this approach is that light only flows forwards in the structure as it progresses through successive layers. This forward-only approach directly avoids forward and backward interference within the structure, hence making it relatively insensitive to precise layer thicknesses or spacings, a potentially significant advantage for manufacturing. Because it can exploit standardized layers, such individual layers could be made and functionally tested separately in advance and then stacked, further simplifying manufacture and increasing yield. Hence there is arguably a path to very complex, functional, and manufacturable multilayered optical structures, avoiding effective design volume limitations of approaches such as inverse design and the challenging fabrication of vertically and horizontally precise stacks of very thin, complex layers.

The architecture exploits layers of $2\times 2$ interferometer blocks, and it could be applied immediately to planar silicon photonic circuits, offering a new architecture suited to wide circuits of limited architectural depth and hence compatible with the long and thin nature of programmable MZI blocks in such technologies, which restricts the number of successive layers of blocks.

An important attribute of this architecture is that it supports self-configuration. In technologies that support full programmability (e.g., of phase shifters in the blocks), this is an obvious benefit because it means that the device can be set up just using input vectors corresponding to the rows of the matrix to be implemented by the device, without calibration or calculations, and just by successive single-parameter power optimizations. Even if we cannot program the device after manufacture, the self-configuration property guarantees that, if we know what linear function (and hence matrix) we want the device to implement, we have a simple progressive technique for designing the settings of the structure without any iterations or optimization procedures; essentially, we can pretend we are self-configuring the structure, simply writing down the resulting settings as we proceed through the nominal self-configuration, block by block.

Unlike design approaches based on global optimization, where it may not be clear in advance if a desired design is possible given starting materials and sizes, we can also be quite clear just what mathematical functions can be implemented by this layered approach. Mathematically, for a structure of $m$ successive layers of $2\times 2$ blocks, it can implement arbitrary unitary (or, technically, semi-unitary) banded-diagonal matrices with bandwidth (number of the possibly non-zero elements in the rows) of $2m-2$, with rows grouped in pairs (so technically a set of $2\times(2m-2)$ blocks on the matrix diagonal). We can also readily deduce the overlapping nonlocality [9] $C$ that such a layered structure can support (e.g., a 4 layer structure can support $C$ up to 4 – see Supplemental document section S1).

We have shown some examples of useful functions with such an approach. Even a single layer of such blocks could allow a camera that measures both the conventional intensity image as well as the phase tilt of the incident light. Other known highly flexible architectures such as a binary tree, which can route any $2m$ element vector to a single output, can be embedded in this approach, allowing an array of such functions. Sophisticated signal processing functions such as filters based on Daubechies wavelet coefficients can be efficiently implemented, allowing parallel convolution of such a filtering function or kernel over a whole array. By illuminating at an angle, complex spectral functions can be implemented. Even a single block can be configured to measure the wavelength of an incoming monochromatic light beam.

Despite these many advantages and capabilities, significant challenges remain in applying this concept to layered metastructures. The first and arguably most important challenge is to design a metastructure that can implement an array of $2\times 2$ beamsplitters. While there are approaches in principle to this, it is not clear if any existing design accomplishes this. Second, we need a metastructure layer of uniform thickness that can nonetheless be designed potentially to implement any of a range of phase delays in transmission (though generating such phase delays is a major strength of metasurfaces, with many possible approaches [30]). Third, all these layered components should have negligible back-reflection.

We note too that this approach requires that we take a "pixelated" approach to the optical system; we must divide the input field in this way so that we can prepare it for such an architecture of connected blocks. Though many approaches to this are possible, such as arrays of lenslets at the inputs, this is a departure from typical metastructure design. Such a pixelation requirement and the implementation of the $2 \times 2$ functionality of the blocks might also tend to limit the minimum effective pixel size. Also, though we have argued above that the manufacture could be relatively simplified by using separately manufactured and standardized layers (such as arrays of $2 \times 2$ beamsplitters), we would still need to devise a practical technique for stacking multiple such layers. We have, however, sketched one approach that would allow the use of just one such layer of $2 \times 2$ beamsplitters together with a spatial light modulator and an optional image sensor that would allow a fully programmable and even self-configuring multiple layer structure for a linear array of inputs.

In this introduction of this architecture and associated algorithms, I have illustrated a few example designs and introduced various key concepts. There are, however, many open questions for further research, on topics that we have only been able to introduce briefly here.

1) With extensions as in the architecture in Fig. 9, this approach can implement non-unitary as well as unitary operations. Many important operations, such as derivatives, for example, require non-unitary operations [17].

2) Most of the discussion has explicitly considered only linear arrays of inputs and outputs. Many two-dimensional arrays of blocks could also be considered, of which only one simple example has been presented here (Fig. 10).

3) When made with repetitive collections of blocks, this approach can formally implement a kind of "forward" photonic crystal (discussed in Supplemental document section S7). This approach may enable the simple design of a variety of such periodic structures. Possible connections to and extensions of other photonic crystal work based on constructive lateral couplings over finite distances [31–34] are intriguing for future work.

More subtly and speculatively, we can also argue that, if we can design some structure that performs a function of interest using this approach, it proves that the function is possible in principle, which might then give us the confidence to approach implementing the function by some other more highly optimized design approach, with this approach also giving us some hints as to how to do that.

## 9. Conclusions

I have argued that there are architectures for layered photonic structures that can allow the simple, progressive design of complex and highly functional optical structures; the design itself can be extended to high complexity because of the simplicity of the progressive calculations to design or simulate its behavior, which only require simple matrix multiplications. With some technological approaches, the desired functions could also be implemented by self-configuration; the design algorithm supports this explicitly. This simplicity does require that we pixelate the system and that we approach the design and the structures themselves in different ways from typical metastructures; this approach is based on a succession of relatively standard layers, though, importantly, existing metastructure approaches, such as inverse design, could themselves be very useful in the design of such standard layers. Hopefully, this approach encourages us that there is a broad range of highly functional complex multilayered optics that we will be able both to design and to manufacture.

**Data availability.** No data were generated or analyzed in the presented research.

**Disclosures.** The author declares no conflicts of interest related to this article.

# Moving forward with multilayer metastructures - simple design of complex optics: Supplemental document

**DAVID A. B. MILLER**

*Ginzton Laboratory, Stanford University, 348 Via Pueblo Mall, Stanford CA 94305 USA*
*dabm@stanford.edu*

### S1 Overlapping nonlocality and meshes

The concept of overlapping nonlocality (Ref. [9] of the main text) is the idea that in the flow of the optical field from the input surface to the output surface in some linear optical device, some number $C_{RL}$ of independent channels of information must flow from the right of the input surface to the left of the output surface, and similarly some number $C_{LR}$ of channels must flow from the left of the input surface to the right of the output surface. The overlapping nonlocality $C$ of this optical system is then the sum

$$C = C_{LR} + C_{RL} \tag{21}$$

This number of channels can be thought of as the number that must be able to flow through a "transverse aperture" formed by dividing the input and output surfaces into their "left" and "right" halves, as in Fig. S1(a).

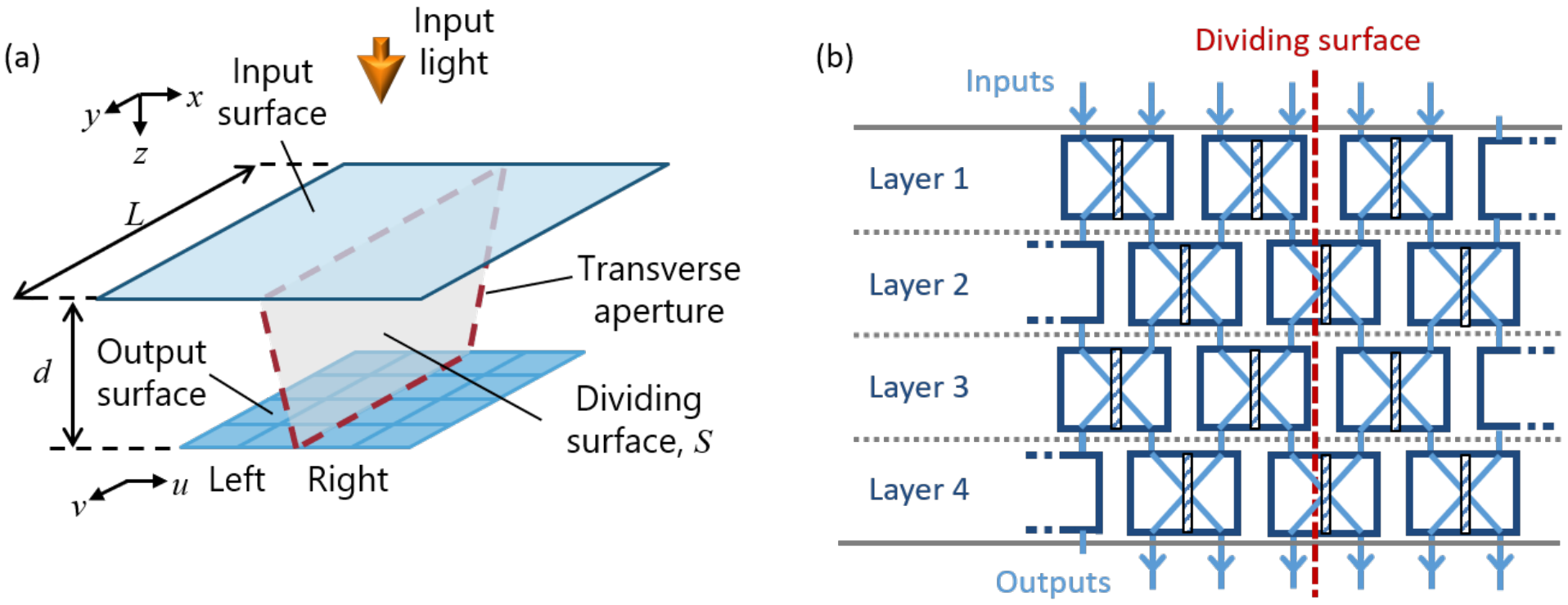

Fig. S1. Overlapping nonlocality. In (a), we divide a device mapping fields on the input surface to fields on the output surface using a dividing surface though both the input and output surfaces, defining a transverse aperture. The aperture must be large enough to support the overlapping nonlocality, the number of independent optical channels that must flow from the input on the left to the output on the right and from the input on right to the output on the left, adding the total to get the overlapping nonlocality C. (b) A portion of a four-layer mesh, with $2\times2$ blocks shown as beamsplitters (with additional phase shifters omitted for simplicity), showing that a dividing surface from the input to the output passes through 4 waveguides or paths, so supporting an overlapping nonlocality of up to 4.

One might think that the optics in the transverse aperture might just need to support one or other of these numbers of channels. For example, the cross-sectional area to support a number of waveguide modes flowing from inputs on the right to outputs on the left would be just the same as the cross-sectional area to support the same number of waveguide modes flowing from inputs on the left to outputs on the right. However, if our optics is reciprocal, then corresponding backward channels must exist. For example, for each such waveguide mode flowing from inputs on the left to outputs on the right, the backwards version of that waveguide mode, flowing backwards from the outputs on the right to the inputs on the left must also exist, and we note that is a channel that is flowing from right to left through the transverse aperture; the number of such backwards right-output-to-left-input channels is, by this reciprocity, also $C_{LR}$ . So, the total number of channels that must exist for flowing from right to left through the transverse aperture is the sum as in Eq. (1). We get the same answer for the total number of channels that must exist for flowing from left to right through the transverse aperture.

To understand what overlapping nonlocality is required from our optics, we can imagine that we start with some matrix $D$ that represents the optical system we want to make. This is a matrix mapping input degrees of freedom – for example, fields at some discrete number of points on the input surfaces (or such samples of the input field) – to output degrees of freedom – for example pixels on some output sensor. We can immediately deduce just what overlapping nonlocality is required in the optics. We divide the matrix $D$ into quadrants; we divide the columns of the matrix with a (vertical) line corresponding to the position of the transverse aperture on the input surface, and we divide the rows with a (horizontal) line corresponding to the position of the transverse aperture on the output surface. (See Ref. [9] Fig. S7 for an example of such a division of a matrix.) We can then perform singular value decomposition on, say, the right-inputs-to-left-outputs quadrant; the number of significant singular values (i.e., with magnitudes significantly larger than zero) tells us the number $C_{RL}$ right-inputs-to-left-outputs channels we require. Similarly, considering the left-inputs-to-right-outputs quadrant, we deduce $C_{LR}$ . Hence we can deduce the minimum required overlapping nonlocality we need in our optical system from Eq. (1).

Notably, many optical operations require that the overlapping nonlocality is a significant number. In turn, this means that the operation cannot be performed just between the input and output surfaces of a single thin optical element because there are insufficient transverse channels available inside the element.

In the case of the layered structure discussed here, such as in the example in Fig. 1(c) or Fig. 4 of the main text, it is straightforward to understand just what overlapping nonlocality the structure can provide; we simply have to count the waveguides or beam paths that cross a corresponding "transverse aperture" surface, counting those that go from left to right and those that go from right to left, and adding the two to get the available overlapping nonlocality *C* of the structure.

In the case of the $2\times 2$ blocks, we need to count the input and output waveguides or paths that must be crossed by a such dividing surface or line. This is easiest to see if we view the blocks as (controllable) beamsplitters as in Fig. S1(b) (with additional phase shifters, not shown) because then we do not get confused by "cutting through" the waveguides inside the MZI, which are merely there to provide the overall beamsplitting function, not to route channels from left to right or *vice versa*. Equivalently, we can view this more topologically, allowing ourselves to distort the dividing surface so that it routes round the internal waveguides; the overlapping nonlocality can then be viewed as the minimum number of paths that must be crossed by any line starting and ending at the relevant input and output positions.

We see in the example in Fig. S1(b), which corresponds to a section of the 4-layer mesh as in Fig. 1(c) and Fig. 4 in the main text, that the dividing surface cuts through exactly 4 waveguides – two going from left to right and two going from right to left – leading to this structure being able to support an overlapping nonlocality of up to $C = 4$ .

One particularly important point for our proposed architectures is that they require devices (such as an MZI) that can function as $2\times 2$ splitters, which require an overlapping nonlocality of 2. This is obvious just from such $2\times 2$ beamsplitters in Fig. S1; any dividing surface through such a block necessarily crosses two paths. (Note that for a "Y" beamsplitter, such a surface need only cross one path.) In general, this means we should not expect to be able to fabricate a $2\times 2$ beamsplitter just from one thin (e.g., subwavelength) layer.

### S2 Matrix for a 2x2 self-configuring block

It is straightforward to deduce the matrix represented by a $2\times 2$ MZI block once we have self-configured it to route all the input power to, say, the upper ("Right") output as in Fig. 1(a) of the main text. We give this algebra explicitly here.

To keep the algebra as simple as possible, as in Fig. 1(a) of the main text, we use phase shifter pairs differentially; that is, the input phase shifters $\phi_T$ and $\phi_L$ are run as a differential pair, so with relative phase shift values $\phi_L = -\phi_T$, and a relative phase shift

$$\Delta\phi = \phi_T - \phi_L \tag{22}$$

Similarly, the internal phase shifters are also run as a differential pair with phase shift values $\phi_W = -\phi_P$, with relative phase shift

$$\Delta\theta = \theta_P - \theta_W \tag{23}$$

and with no "common-mode" phase shift from either pair. (We can always add the effect of any such common-mode phase shift $\phi_{CM}$ as an overall multiplying factor $\exp(i\phi_{CM})$ later if needed.)

We presume the input vector of mode amplitudes is

$$|e_{in}\rangle = \begin{bmatrix} a \\ b \end{bmatrix} \tag{24}$$

with $a$ and $b$ being the electric field amplitudes in the "Top" and "Left" inputs respectively. The resulting corresponding output vector $|e_{out}\rangle$ of amplitudes in the "Right" and "Bottom" outputs from the MZI block is the result of passing this input through the MZI. We describe this linear operation by the total matrix $\mathsf{M}_T$, so we have

$$|e_{out}\rangle = \mathsf{M}_T |e_{in}\rangle \tag{25}$$

Our task here is to find the matrix $\mathsf{M}_T$ if we self-configure this MZI block to route all the power to the upper ("Right") output. (We could similarly find the corresponding matrix if instead we chose to route to the lower ("Bottom") output.)

We can describe the overall matrix $\mathsf{M}_T$ in terms of a matrix $\mathsf{M}_\phi$ corresponding to passing through the input $\phi$ phase shifters, which is just the diagonal matrix

$$\mathsf{M}_\phi(\Delta\phi) = \begin{bmatrix} p & 0 \\ 0 & p^* \end{bmatrix} \tag{26}$$

where

$$p = \exp(i\Delta\phi/2) \tag{27}$$

(and $p^* = \exp(-i\Delta\phi/2)$) and the matrix $\mathsf{M}_S$ corresponding to passing through the rest of the MZI block (with no output phase shifters $\theta_R$ or $\theta_B$), which is derived in detail in Ref. [13] of the main text. Presuming 50:50 beamsplitters this is given by

$$\mathsf{M}_S(\Delta\theta) = \begin{bmatrix} s & c \\ c & -s \end{bmatrix} \tag{28}$$

where

$$s = \sin(\Delta\theta/2) \text{ and } c = \cos(\Delta\theta/2) \tag{29}$$

$\mathsf{M}_T$ is then the product

$$\mathsf{M}_T = \mathsf{M}_S\mathsf{M}_\phi = \begin{bmatrix} s & c \\ c & -s \end{bmatrix}\begin{bmatrix} p & 0 \\ 0 & p^* \end{bmatrix} = \begin{bmatrix} ps & p^*c \\ pc & -p^*s \end{bmatrix} \tag{30}$$

or, explicitly,

$$\mathsf{M}_T(\Delta\phi,\Delta\theta) = \begin{bmatrix} \exp\left(i\frac{\Delta\phi}{2}\right)\sin\left(\frac{\Delta\theta}{2}\right) & \exp\left(-i\frac{\Delta\phi}{2}\right)\cos\left(\frac{\Delta\theta}{2}\right) \\ \exp\left(i\frac{\Delta\phi}{2}\right)\cos\left(\frac{\Delta\theta}{2}\right) & -\exp\left(-i\frac{\Delta\phi}{2}\right)\sin\left(\frac{\Delta\theta}{2}\right) \end{bmatrix} \tag{31}$$

Hence, with the input vector of amplitudes $|e_{in}\rangle$ as in Eq.(4), the output vector $|e_{out}\rangle$ is

$$|e_{out}\rangle = \begin{bmatrix} aps + bp^*c \\ apc - bp^*s \end{bmatrix} \tag{32}$$

Suppose that we self-configure this MZI block, minimizing the “Bottom” output power, first by adjusting the input differential phase shift $\Delta\phi$ and then the internal differential phase shift $\Delta\theta$ to minimize the “Bottom” output power to zero. Then, the “Bottom” field amplitude must be zero, so

$$apc = bp^*s \tag{33}$$

Writing

$$a = |a|\exp(i\alpha),\ b = |b|\exp(i\beta) \tag{34}$$

from Eqs. (7) and (13)we therefore have

$$\frac{p}{p^*} = \exp(i\Delta\phi) = \frac{b}{a}\frac{s}{c} = \frac{|b|}{|a|}\exp\left[i(\beta-\alpha)\right]\tan\left(\frac{\Delta\theta}{2}\right) \tag{35}$$

Since $\tan(\Delta\theta/2)$ is a real number, we therefore have

$$\Delta\phi = \beta - \alpha \tag{36}$$

and

$$\Delta\theta = 2\tan^{-1}\left(|a/b|\right) \tag{37}$$

So, if we self-configure an MZI block using an input $|e_{in}\rangle$ as in Eq. (4) so as to put all the output in the “Right” output port, using the forms (14) for the input amplitudes, then the resulting matrix for the block is given by $\mathsf{M}_T$ as in Eq. (11), with the resulting phase shifter settings and matrix parameters $\Delta\phi$ given by Eq. (16) and $\Delta\theta$ given by Eq. (17). Note these mathematical results are simple arithmetic calculations from the input used to set the block by self-configuration.

Note, incidentally, that this analysis clarifies that we only need a range of a total of $\pi$ for the relative phase shift $\Delta\theta$. The argument in the inverse tangent in Eq. (17) is always positive, which means that the resulting $\Delta\theta$ only runs between 0 and $\pi$.

Note also that this analysis proves mathematically another point that is obvious physically: for self-configuration, we require that we have controllable relative phase shift $\Delta\phi$ on the inputs to the block. If we put phase shifters only on the outputs of the block, then the matrix multiplication is reversed to give a matrix

$$\tilde{\mathsf{M}}_{TOT} = \mathsf{M}_{\phi}\mathsf{M}_{S} = \begin{bmatrix} ps & pc \\ p^{*}c & -p^{*}s \end{bmatrix} \tag{38}$$

With the input vector $|e_{in}\rangle$ as in Eq. (4)as before, the "Bottom" output amplitude is $ap^{*}c - bp^{*}s$. For this to be zero, because the $p^{*}$ is a common factor, we deduce $a/b = \tan(\Delta\theta/2)$, which is a real number. So, we can only get such cancellation if the input vector elements are in phase with one another, which in general they are not. Hence, without input phase shift control, we cannot in general self-configure the MZI to null out the "Bottom" output.

For completeness, we can repeat the calculation for the case where we want all the output in the "Bottom" output port. Starting from Eq. (12), to route all the power to the "Bottom" output by minimizing the "Right" field amplitude to be zero, we have, instead of Eq. (13),

$$aps = -bp^{*}c \tag{39}$$

So

$$\begin{aligned} \frac{p^{*}}{p} = \exp(-i\Delta\phi) = -\frac{a}{b}\frac{s}{c} = -\frac{|a|}{|b|}\exp\left[i(\alpha-\beta)\right]\tan\left(\frac{\Delta\theta}{2}\right) \\ = \frac{|a|}{|b|}\exp\left[i(\alpha-\beta-\pi)\right]\tan\left(\frac{\Delta\theta}{2}\right) \end{aligned} \tag{40}$$

Here we have chosen to replace the minus sign with $\exp(-i\pi)$. So, again, since $\tan(\Delta\theta/2)$ is a real number, we have

$$\Delta\phi = \beta - \alpha + \pi \tag{41}$$

and

$$\Delta\theta = \tan^{-1}\left(|b/a|\right) \tag{42}$$

## S3 Use of an MZI block to measure wavelength

In the main text, we discussed the use of a single MZI block to measure the phase difference between the two inputs to the block. Another related use of individual blocks is to measure wavelength of a monochromatic beam. In this case, we also presume the beam is incident on the block inputs at some angle $\alpha$ to the vertical. Then different wavelengths will have different phase differences between the block inputs.

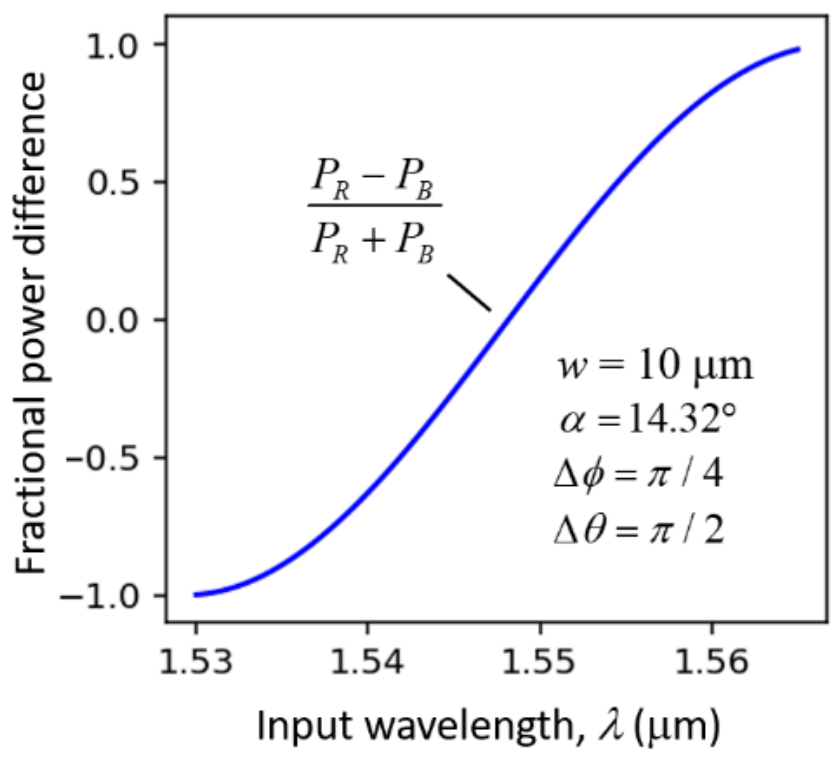

Fig. S2. Use of an MZI block to measure the wavelength of a beam incident at an angle of 14.45º. The two inputs to the block are spaced by 10 mm. Block settings are $\Delta\phi = \pi / 4$ and $\Delta\theta = \pi / 2$ .

As an example, we choose settings of $\Delta\phi = \pi / 4$ and $\Delta\theta = \pi / 2$ for the block and calculate the resulting matrix $\mathsf{M}_T$ from Eq. (6) of the main text. Again, we choose $w = 10$ microns and now choose an input angle of 14.32º, which gives a useful coverage for the telecommunications C band. The phase difference between the two inputs $\phi_a = (2\pi / \lambda) w \sin\alpha$ as in Eq. (10) of the main text and the calculated fractional power difference between the two outputs is shown in Fig. S2, showing that such a simple block could also be used for wavelength measurement.

### S4 Mesh matrices as unitary operators

There is a subtle technical point as to whether our mesh of $2 \times 2$ blocks is unitary. Though the rows of the matrix $\mathsf{U}_M$ describing the mesh behavior are orthogonal and normalized, $\mathsf{U}_M$ may not be regarded technically as a unitary matrix because it has different numbers of rows and columns; such matrices can be called semi- or partially unitary. However, as an operator that maps from one Hilbert space (the "input" space) to another (the "output" space), it is unitary in the general sense because those Hilbert spaces, spanned respectively by the Hermitian adjoints of the rows ("input" space) and the columns ("output" space) of the matrix, have the same dimensionality (here, 8). If we were to regard the input space as being 14-dimensional in our example (because there are 14 inputs) rather than 8 dimensional, so capable to representing any 14 element vectors, then indeed the operation is not unitary; in particular, physically, there could be power routed out of the other (un-numbered and unconnected) "drop" ports in the $2 \times 2$ blocks at the left and right of the structure (such as the left port of block B11), which would make the system physically non-unitary because it has loss. With this understanding, though, we can essentially view these mesh matrices as unitary.

### S5 Self-configuring algorithm for setting the trapezoidal block layer mesh for arbitrary orthogonal inputs

Here we give a full version of the configuration algorithm for the mesh as in Fig. 4 in the main text for the general case; so, we presume no particular space-invariance, with each successive configuration Vector is as independent as it can be if it is to be routed to a different output. The algorithm corresponding to the matrix $\mathsf{U}_M$ as in Eq. (14) of the main text, is as follows.

Layer 1 settings
  Shine in only Vector 1
    Adjust block B11 to set the output power in its left output port to zero
    Adjust block B14 to set the output power in its right output port to zero.

(Note: these steps ensure both that power is flowing "rightwards" from B11 towards Output 1 and "leftwards" from B14 towards Output 1)

Shine in only Vector 2

Adjust block B12 to set the output power in its left output port to zero

Adjust block B15 to set the output power in its right output port to zero.

Shine in only Vector 3

Adjust block B13 to set the output power in its left output port to zero

Adjust block B16 to set the output power in its right output port to zero

Shine in only Vector 4

(We do not need to adjust block B13 to set the output power in its left output port to zero because B13 has already been set in the Vector 1 steps above.)

Adjust block B17 to set the output power in its right output port to zero

Layer 2 settings

Shine in only Vector 1

Adjust block B21 to set the output power in its left output port to zero

Adjust block B23 to set the output power in its right output port to zero.

(Note: these steps ensure both that power is flowing "rightwards" from B21 towards Output 1 and "leftwards" from B23 towards Output 1)

Shine in only Vector 2

Adjust block B22 to set the output power in its left output port to zero

Adjust block B24 to set the output power in its right output port to zero

Shine in only Vector 3

(We do not need to adjust block B23 to set the output power in its left output port to zero because B23 has already been set in the Vector 1 steps above.)

Adjust block B25 to set the output power in its right output to zero

Shine in only Vector 4

(We do not need to adjust block B24 to set the output power in its left output port to zero because B24 has already been set in the Vector 2 steps above.)

Adjust block B26 to set the output power in its right output to zero

Layer 3 settings

Shine in only Vector 1

Adjust block B31 to set the output power in its left output port to zero

Adjust block B32 to set the output power in its right output port to zero.

Shine in only Vector 2

Adjust block B33 to set the output power in its right output port to zero

Shine in only Vector 3

Adjust block B34 to set the output power in its right output port to zero

Shine in only Vector 4

Adjust block B35 to set the output power in its right output port to zero

Layer 4 settings

Shine in only Vector 1

Adjust block B41 to set the power in its right output port to zero

Shine in only Vector 2

Adjust block B42 to set the power in its right output port to zero

Shine in only Vector 3

Adjust block B43 to set the power in its right output port to zero

Shine in only Vector 4

Adjust block B44 to set the power in its right output port to zero

Having set this first trapezoid of blocks this way, we then proceed with a similar but slightly simpler algorithm with successive vectors to set up any remaining blocks in a wider array. In this case, we could continue with a Vector 5, which is shifted over by one more block to the right and which must be orthogonal to Vectors 2, 3, and 4 (it is automatically orthogonal to Vector 1 because it does not overlap with it). The algorithm for Vector 5 is

Shining in only Vector 5

Adjust block B18 to set the output power in its right output port to zero
Adjust block B27 to set the output power in its right output port to zero
Adjust block B36 to set the output power in its right output port to zero
Adjust block B45 to set the output power in its right output port to zero

Note that this algorithm for Vector 5 is like that for a diagonal line self-configuring layer, just with the constraints on Vector 5 being orthogonal to the preceding 3 Vectors (here Vectors 2, 3, and 4).

We can continue with further Vectors, each shifted to the right by one block, and each orthogonal to the preceding 3 Vectors. We see then that, from Vector 4 onwards, the mesh is being configured as a set of successive diagonal line self-configuring layers: blocks B17, B26, B35, and B44 for Vector 4; blocks B18, B27, B36, and B45 for Vector 5; and so on. In this way, we can set an arbitrarily wide set of such layers of blocks with these various 8-element shifted input Vectors of amplitudes.

For all these algorithms, note that there are additional Vectors $1'$, $2'$, $3'$, $4'$, $5'$, and so on (not shown), of the same size and positions as the Vectors 1, 2, 3, 4, 5, and so on, that will be routed to the other outputs of the last block layer of blocks, i.e., outputs 2, 4, 6, 8, 10, and so on. These complementary vectors are completely determined by the mesh as set by the Vectors 1, 2, 3, 4, 5, and so on. They have no additional degrees of freedom (other than possibly their overall phase). To understand what they are, and to calculate them, once we have set the mesh with Vectors 1, 2, 3, 4, 5, and so on, we now imagine shining light back into these outputs 2, 4, 6, 8, 10, and so on, one at a time. The vector of amplitudes that emerges backwards from the corresponding set of inputs in each case will be the phase conjugate of the corresponding Vector $1'$, $2'$, $3'$, $4'$, $5'$, and so on.

We have illustrated this process of configuration of these block layers of blocks for the example case of 4 such block layers. The extension of other numbers of layers is straightforward. For $m$ such block layers, we will be using vectors with $2m$ elements for the configuration or calculation, with each successive vector orthogonal to the preceding $m-1$ configuration Vectors.

### S6 Algorithm based only on self-configuring diagonal lines

If we are prepared to discard a few inputs and outputs at the "left" of the mesh, we can self-configure the mesh just using each configuration vector once. The algorithm is illustrated in Fig. S3.

In this algorithm, we presume that we have a set of configuration Vectors, each $2m$ inputs long, and each orthogonal to every other such Vector, just as in the algorithm above in Section S5. However, now we block off the first $m-2$ inputs to the mesh (i.e., Inputs 1 through 6 in our example). Also, though we may use photodetectors in the outputs of the first $m-1$ output blocks – Outputs 1, 3, and 5 in our example – we otherwise discard the Outputs from these blocks (Outputs 1 through 6 in our example).

The algorithm then proceeds as follows:

Shine in only Vector 1

Configure the first diagonal line (which starts with Inputs 7 and 8) to route all the power to the left output at the end of the diagonal line (here Output 1)

Proceed similarly for all diagonal lines the mesh, working progressively to the right, with the *n*th such diagonal line configured with the *n*th Vector.

Note in this algorithm that we may set the blocks marked with a cross to "cross" state to start with. (Indeed, we could choose to set all blocks to "cross" state to start with, which makes the use of output photodetectors easier as the requisite signals are already routed to them.) Even if we do not set them to "cross", an algorithm to maximize the power from these first $m-1$ output blocks will end up setting such blocks to "cross" anyway.

Note that, when shining in a given Vector *n* and maximizing the power from the bottom left output of the corresponding diagonal line *n*, we can simultaneously be shining any power into the inputs to the right of that Vector because none of it can be routed to that bottom left port. This could be particularly convenient if setting up a periodic mesh as in Section S8 below because we could simply shine in a complete set of replicated versions of that Vector into the successive unit cells of the structure. (So, Vector 5 is the same as Vector 1, Vector 6 is the same as Vector 2, and so on.) Then the entire periodic structure can be set up with just *m* such repetitive input vectors. (Note that, because of the way we have constructed this algorithm, it would be more convenient to view the unit cell of such a crystal structure as a parallelogram skewed to the left rather than skewed to the right as in Fig. S3. Either one is a possible choice of unit cell for this structure.)

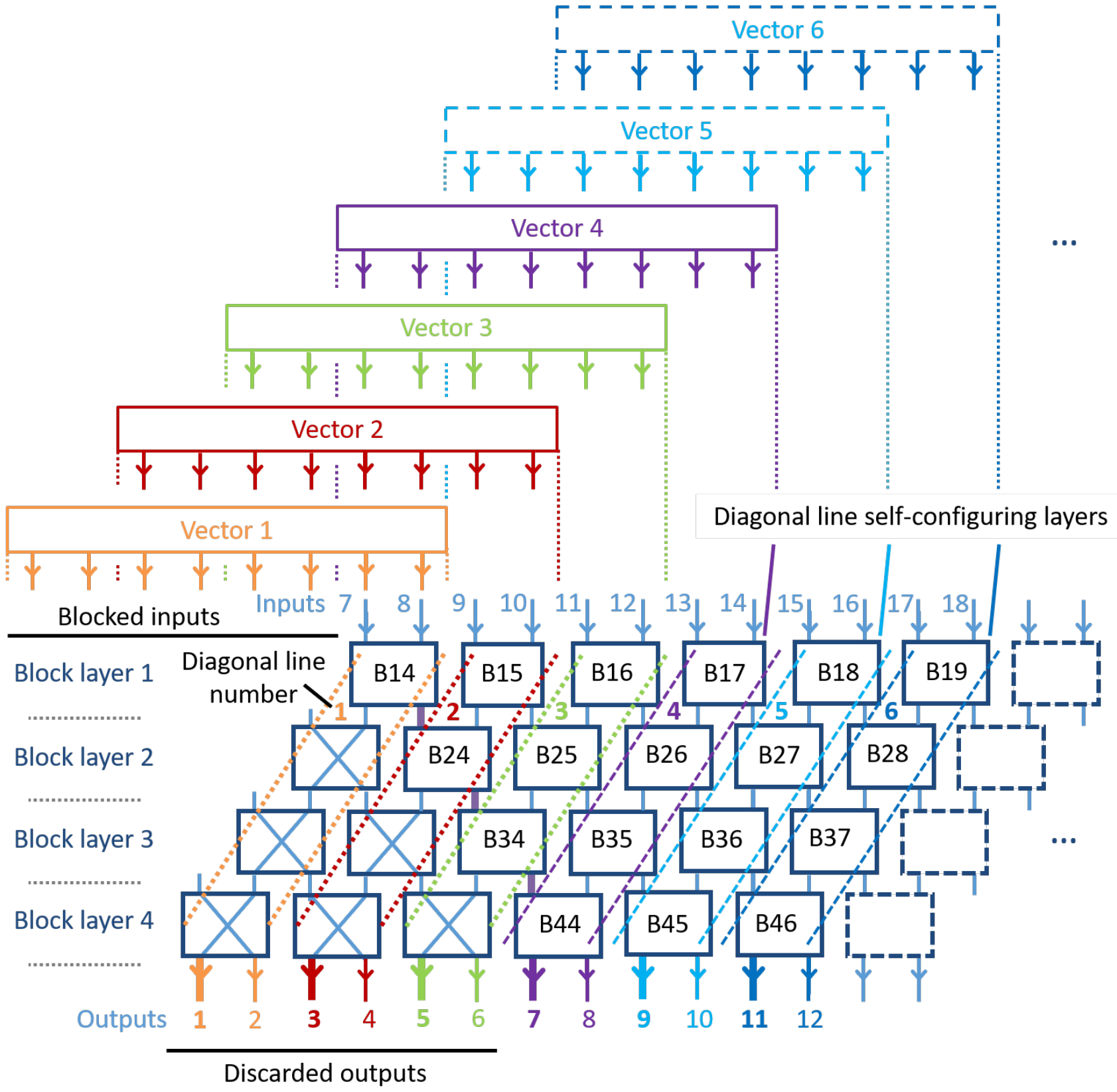

Fig. S3. Architecture for configuring the mesh using just self-configuration of a succession of diagonal lines, one per (orthogonal) input Vector, to maximize output power at the bottom left output of the diagonal line when illuminated with the corresponding Vector. In this architecture, Inputs 1 through 6 are blocked, and the signals from Outputs 1 through 6 are discarded except for their use during configuration. The mesh is then fully configured for Outputs 7 and higher and for Vectors 4 and higher.

## S7 Periodic trapezoidal architectures as photonic crystals

The analysis of the trapezoidal architecture in the main text and in sections S5 and S6 above is general; for a mesh with *m* layers of $2\times 2$ blocks; every one of the $2m$ element configuration Vectors 1, 2, 3, … (or the corresponding matrix rows with $4m-2$ non-zero elements $|\psi_1\rangle$, $|\psi_3\rangle$, $|\psi_7\rangle$ ... ) can be chosen quite arbitrarily to be different from any other as long as each such Vector or matrix row is orthogonal to the others; in this case potentially each $2\times 2$ block could be configured differently from every other block. As mentioned there, the trapezoidal architecture can be arbitrarily extended sideways with more diagonal lines, each with its corresponding additional configuration Vector.

An interesting subset of such universal trapezoidal architectures is structures that are "spatially invariant" – that is, they look the same if translated sideways by some specific amount. In this case, we would have "unit cells" of blocks; the blocks within each unit cell might all be configured differently, but the configuration of every unit cell of blocks would be the same.

With such unit cells, the structure becomes technically a kind of photonic crystal, though with the unusual characteristic that the light only flows in the forward direction; we can refer to these as 'forward' photonic crystals. This characteristic makes their modeling particularly simple; calculating their behavior merely requires matrix multiplications. We examine some classes of these photonic crystals below.

### *Mesh with transverse periodicity of m blocks for an m layer mesh*

Consider a mesh of $m$ layers of blocks that is also periodic in the transverse direction (i.e., within each layer) where the period or "unit cell size" in the transverse direction is also $m$ blocks, or equivalently a periodicity of

$$q = 2m \tag{43}$$

inputs. We construct such a mesh physically by adding arbitrary numbers of "diagonal lines" of blocks as in Fig. 1(c) or Fig. 4 of the main text. As we discuss below, this mesh can be viewed as being made up from parallelogram unit cells stacked side by side (Fig. S4).

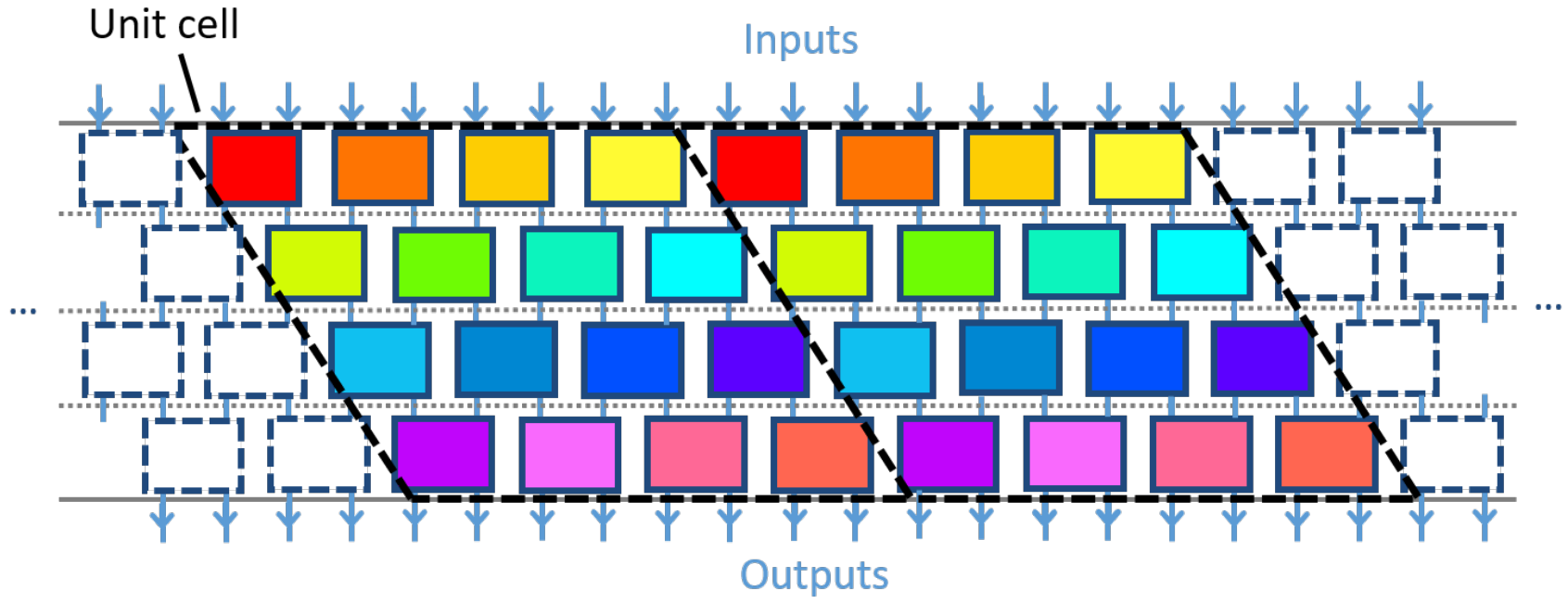

Fig. S4. Sketch of 2 unit cells of a forward photonic crystal structure made from 4 layers of blocks in a mesh. Each of the 16 blocks in a unit cell, shown in different colors, may be set differently in the design or configuration of the mesh, with these block settings replicated in other unit cells.

This approach allows the mesh to have $2m$ different functions that repeat with periodicity of $m$ blocks. This mesh can be set up with $m$ Vectors just as for the arbitrary mesh above, with those $m$ Vectors shifted right by $m$ blocks (or $q = 2m$ inputs) to set up each successive set of $m$ blocks; equivalently, we simply replicate the settings for one such unit cell into the others.

The matrix for this mesh is made up with a set of blocks along the diagonal, with zeros elsewhere. Each such matrix block, which has dimensions $2m\times(4m-2)$, is in the form of $\mathrm{U}_M$ as in Eq. (14) of the main text; so in our example with $m = 4$, these blocks are $8\times 14$. The resulting matrix can be written for this example as

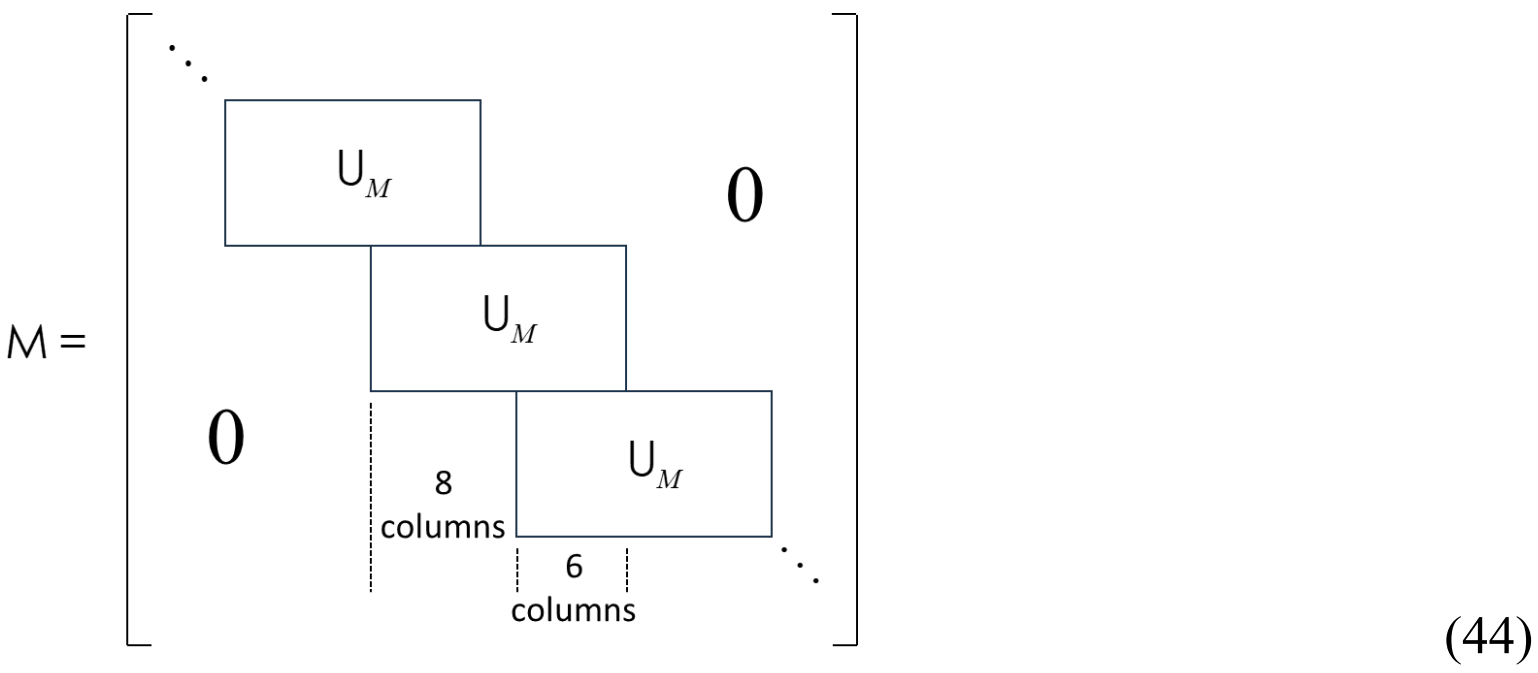

(44)

Note that these blocks are staggered as shown, with each block shifted down and to the right by $2m$ rows and $2m$ columns. If we ignore the grouping into the larger $\mathsf{U}_M$ blocks, the resulting matrix is a block diagonal matrix with blocks that each are 2 rows by $4m-2$ (here 14) columns, just as we see inside $\mathsf{U}_M$ as in Eq. (14) of the main text.

Now we proceed to construct a way of looking at this system like a one-dimensional photonic crystal viewed in transmission from inputs to outputs. To analyze this system, we start by looking for eigenvectors – overall input vectors $|\psi\rangle$ that extend over the whole larger crystal and that map to output versions $|\psi_o\rangle$ of themselves just multiplied by an eigenvalue $\mu$, i.e.,

$$|\psi_o\rangle = \mu|\psi\rangle \tag{45}$$

Note this eigenproblem is slightly different from many typical physical eigenproblems, such as in resonators, in which the mapping is from a space back onto itself; here the mapping is from the input space to the output space, but we are still interested in functions that reproduce themselves within some factor (the eigenvalue $\mu$) in this mapping.

Incidentally, we know immediately that, if such eigenvectors exist, because the mesh is nominally lossless, the eigenvalue must be a unit amplitude complex number, i.e., defining the corresponding "eigenphase" $\eta$, we have

$$\mu = \exp(i\eta) \tag{46}$$

so the interesting behavior of such eigenvectors in passing through the mesh will just be a phase shift $\eta$.

We look for eigenvectors that have some kind of periodicity associated with the periodicity of the mesh "lattice", so with a periodicity of $q = 2m$ inputs (and outputs) as in Eq. (23). $q$ is also the number of matrix rows associated with each period of this photonic crystal lattice. We take the individual elements of the vector $|\psi_o\rangle$ to be the complex numbers $\psi(s)$, where $s$ indexes the mesh outputs; note $s$ is an effective position coordinate, including within the unit cell. We take the transmitted power in such an output $s$ to be

$$P(s) = |\psi(s)|^2 \tag{47}$$

in some convenient units. Now we expect that, when illuminated by such an eigenvector, the transmitted power from the ports will look the same in every period of the mesh. So,

$$|\psi(s+q)|^2 = |\psi(s)|^2 \tag{48}$$

which means that

$$\psi(s+q) = C\psi(s) \tag{49}$$

where $C$ is a unit complex number. Next, we indulge in the usual fallacy of periodic boundary conditions in which we persuade ourselves that the behavior of an infinitely long chain of such periods of the mesh could not really be any different from that of a very long chain joined onto itself in a loop. Taking $S$ periods in that loop, we then also argue that the resulting amplitudes must be single-valued so these amplitudes round the loop must join onto one another. Hence, we argue that

$$\psi(s+Sq) = C^S \psi(s) = \psi(s) \tag{50}$$

So,

$$C^S = 1 \tag{51}$$

and so $C$ is one of the $S$ roots of unity. (We are, of course, deriving a form of the Bloch theorem.) In keeping with common ways of writing this in solid state physics, we can choose to write

$$C = \exp(i\kappa q) \tag{52}$$

where the "wavevector" $\kappa$ (here, a dimensionless number) is any of the elements

$$\kappa = \frac{2\pi p}{Sq}; \;\; p = 0, \pm 1, \pm 2, \ldots \pm \frac{S}{2} \tag{53}$$

(where technically we have to drop one or other of the end points $+S/2$ or $-S/2$ to get the correct number $S$ of distinct values). From Eqs. (29) and (32) we therefore have one form of the Bloch theorem

$$\psi_\kappa(s+q) = \exp(i\kappa q)\psi_\kappa(s) \tag{54}$$

where we are now explicitly writing the $\psi$ for a given $\kappa$ by subscripting it. Defining

$$u_\kappa(s) = \psi_\kappa(s)\exp(-i\kappa s) \tag{55}$$

we have the more common form of the Bloch theorem

$$\psi_\kappa(s) = u_\kappa(s)\exp(i\kappa s) \tag{56}$$

where

$$u_\kappa(s+q) = u_\kappa(s) \tag{57}$$

So, $u_\kappa(s)$ can be thought of as a "unit cell" function that is the same in every "unit cell" – here every set of $q$ inputs or $m = q/2$ blocks. The overall eigenfunction is then written in the form of Eq. (36), which is this unit cell function multiplied by an envelope complex exponential "wave" $\exp(i\kappa s)$. As usual with the Bloch theorem, we are not solving the full problem yet, but we are saying that any eigen solution of the problem must have the form as in Eq. (36).

The final step in our analysis is to deduce the eigen equation for $u_\kappa(s)$, which will in general be different for each choice of "wavevector" $\kappa$. For this, we presume, without loss of generality, that we are interested in outputs 1 through $q$ (1 through 8 here), as being a representative unit cell block of the crystal. To calculate that output with an input in Bloch form, we note that the fields in all the inputs 1 through $2q-2$ (1 through 14 here) influence the final output fields in this set of $q$ output ports. If the inputs 1 through $q$ (1 through 8 here) have amplitudes $\psi_\kappa(1), \psi_\kappa(2), \ldots, \psi_\kappa(q)$, then, because the input wave is presumed to be in Bloch form as in Eq. (34), the fields in inputs $q+1$ to $2q-2$ will be of the form

$$\psi_\kappa(q+1) = e_\kappa \psi_\kappa(1),\ \psi_\kappa(q+2) = e_\kappa \psi_\kappa(2),\ \ldots,\ \psi_\kappa(2q-2) = e_\kappa \psi_\kappa(q-2)$$

(58)\

where we are using a shorthand

$$e_\kappa = \exp(i\kappa q) \tag{59}$$

So, the matrix multiplication that determines the outputs 1 through $q$ to give the output vector for these $q$ elements is

$$\begin{bmatrix} \psi_{o\kappa}(1) \\ \vdots \\ \psi_{o\kappa}(q) \end{bmatrix} =$$

$$\begin{bmatrix}
u_{1,1} & u_{1,2} & u_{1,3} & u_{1,4} & u_{1,5} & u_{1,6} & u_{1,7} & u_{1,8} & 0 & 0 & 0 & 0 & 0 & 0 \\
u_{2,1} & u_{2,2} & u_{2,3} & u_{2,4} & u_{2,5} & u_{2,6} & u_{2,7} & u_{2,8} & 0 & 0 & 0 & 0 & 0 & 0 \\
0 & 0 & u_{3,3} & u_{3,4} & u_{3,5} & u_{3,6} & u_{3,7} & u_{3,8} & u_{3,9} & u_{3,10} & 0 & 0 & 0 & 0 \\
0 & 0 & u_{4,3} & u_{4,4} & u_{4,5} & u_{4,6} & u_{4,7} & u_{4,8} & u_{4,9} & u_{4,10} & 0 & 0 & 0 & 0 \\
0 & 0 & 0 & 0 & u_{5,5} & u_{5,6} & u_{5,7} & u_{5,8} & u_{5,9} & u_{5,10} & u_{5,11} & u_{5,12} & 0 & 0 \\
0 & 0 & 0 & 0 & u_{6,5} & u_{6,6} & u_{6,7} & u_{6,8} & u_{6,9} & u_{6,10} & u_{6,11} & u_{6,12} & 0 & 0 \\
0 & 0 & 0 & 0 & 0 & 0 & u_{7,7} & u_{7,8} & u_{7,9} & u_{7,10} & u_{7,11} & u_{7,12} & u_{7,13} & u_{7,14} \\
0 & 0 & 0 & 0 & 0 & 0 & u_{8,7} & u_{8,8} & u_{8,9} & u_{8,10} & u_{8,11} & u_{8,12} & u_{8,13} & u_{8,14}
\end{bmatrix}$$

$$\times \begin{bmatrix} \psi_\kappa(1) \\ \vdots \\ \psi_\kappa(q) \\ e_\kappa \psi_\kappa(q+1) \\ \vdots \\ e_\kappa \psi_\kappa(2q-2) \end{bmatrix}$$

(60)

Now we note a mathematical trick. We first move the triangle of matrix elements in the last $2m-2$ columns on the right over to fill in the zeros on the left, multiplying them by $e_\kappa$. (Formally this is equivalent to multiplying the last $2m-2$ columns by $e_\kappa$ and adding them to the first $2m-2$ columns.) Then we remove the last $2m-2$ columns to get a new $q \times q$ matrix. In our example, this matrix becomes

$$\Psi_{C\kappa}=\begin{bmatrix} u_{1,1} & u_{1,2} & u_{1,3} & u_{1,4} & u_{1,5} & u_{1,6} & u_{1,7} & u_{1,8} \\ u_{2,1} & u_{2,2} & u_{2,3} & u_{2,4} & u_{2,5} & u_{2,6} & u_{2,7} & u_{2,8} \\ e_\kappa u_{3,7} & e_\kappa u_{3,8} & u_{3,1} & u_{3,2} & u_{3,3} & u_{3,4} & u_{3,5} & u_{3,6} \\ e_\kappa u_{4,7} & e_\kappa u_{4,8} & u_{4,1} & u_{4,2} & u_{4,3} & u_{4,4} & u_{4,5} & u_{4,6} \\ e_\kappa u_{5,5} & e_\kappa u_{5,6} & e_\kappa u_{5,7} & e_\kappa u_{5,8} & u_{5,1} & u_{5,2} & u_{5,3} & u_{5,4} \\ e_\kappa u_{6,5} & e_\kappa u_{6,6} & e_\kappa u_{6,7} & e_\kappa u_{6,8} & u_{6,1} & u_{6,2} & u_{6,3} & u_{6,4} \\ e_\kappa u_{7,3} & e_\kappa u_{7,4} & e_\kappa u_{7,5} & e_\kappa u_{7,6} & e_\kappa u_{7,7} & e_\kappa u_{7,8} & u_{7,1} & u_{7,2} \\ e_\kappa u_{8,3} & e_\kappa u_{8,4} & e_\kappa u_{8,5} & e_\kappa u_{8,6} & e_\kappa u_{8,7} & e_\kappa u_{8,8} & u_{8,1} & u_{8,2} \end{bmatrix} \quad (61)$$

If we now start with a mathematical input vector with just the $q$ elements corresponding to the unit cell, the resulting matrix-vector multiplication leads to exactly the same output results as in Eq. (40),

$$\begin{bmatrix} \psi_{o\kappa}(1) \\ \vdots \\ \psi_{o\kappa}(q) \end{bmatrix} = \Psi_{C\kappa} \begin{bmatrix} \psi_{\kappa}(1) \\ \vdots \\ \psi_{\kappa}(q) \end{bmatrix} \quad (62)$$

Now we write the various $\psi_k$ and $\psi_{o\kappa}$ using the form as in Eq. (36), giving

$$\begin{bmatrix} u_\kappa(1)\exp(i\kappa) \\ \vdots \\ u_\kappa(q)\exp(i\kappa q) \end{bmatrix} \equiv \mathsf{D}_\kappa \begin{bmatrix} u_\kappa(1) \\ \vdots \\ u_\kappa(q) \end{bmatrix}, \begin{bmatrix} u_{o\kappa}(1)\exp(i\kappa) \\ \vdots \\ u_{o\kappa}(q)\exp(i\kappa q) \end{bmatrix} \equiv \mathsf{D}_\kappa \begin{bmatrix} u_{o\kappa}(1) \\ \vdots \\ u_{o\kappa}(q) \end{bmatrix} \quad (63)$$

where $\mathsf{D}_\kappa$ is the diagonal matrix

$$\mathsf{D}_\kappa = \begin{bmatrix} \exp(i\kappa) & 0 & \cdots & 0 \\ 0 & \exp(i2\kappa) & & 0 \\ \vdots & & \ddots & \vdots \\ 0 & 0 & \cdots & \exp(i\kappa q) \end{bmatrix} \quad (64)$$

Hence, Eq. (42) becomes

$$\mathsf{D}_\kappa \begin{bmatrix} u_{o\kappa}(1) \\ \vdots \\ u_{o\kappa}(q) \end{bmatrix} = \Psi_{C\kappa}\mathsf{D}_\kappa \begin{bmatrix} u_\kappa(1) \\ \vdots \\ u_\kappa(q) \end{bmatrix} \quad (65)$$

Noting that, since $\mathsf{D}_\kappa$ is unitary, its inverse $\mathsf{D}_\kappa^{-1} \equiv \mathsf{D}_\kappa^\dagger$, which is its Hermitian adjoint, and introducing the short-hand notations

$$\mathsf{U}_\kappa \equiv \begin{bmatrix} u_\kappa(1) \\ \vdots \\ u_\kappa(q) \end{bmatrix}, \ \mathsf{U}_{o\kappa} \equiv \begin{bmatrix} u_{o\kappa}(1) \\ \vdots \\ u_{o\kappa}(q) \end{bmatrix} \quad (66)$$

we can rearrange Eq. (45) to obtain

$$\mathsf{U}_{o\kappa} = \mathsf{U}_{C\kappa}\mathsf{U}_\kappa \quad (67)$$

where

$$\mathsf{U}_{C\kappa} = \mathsf{D}_{\kappa}^{\dagger}\Psi_{C\kappa}\mathsf{D}_{\kappa} \tag{68}$$

is now the "unit cell" matrix.

The corresponding eigenequation for the unit cell functions (or vectors) is

$$\mathsf{U}_{C\kappa}\mathsf{u}_{\kappa p} = \mu_{\kappa p}\mathsf{u}_{\kappa p} \equiv \exp\left(i\eta_{\kappa p}\right)\mathsf{u}_{\kappa} \tag{69}$$

where we introduce a "band" index $p$ since we expect in general there will be $q$ different eigensolutions for each choice of $\kappa$, and have indexed all eigenvectors and eigenvalues with their corresponding values of $\kappa$ and $p$.

Note that, in this periodic case, the resulting full matrix for the mesh can be written as a block diagonal matrix with square blocks $\mathsf{U}_{C\kappa}$, as

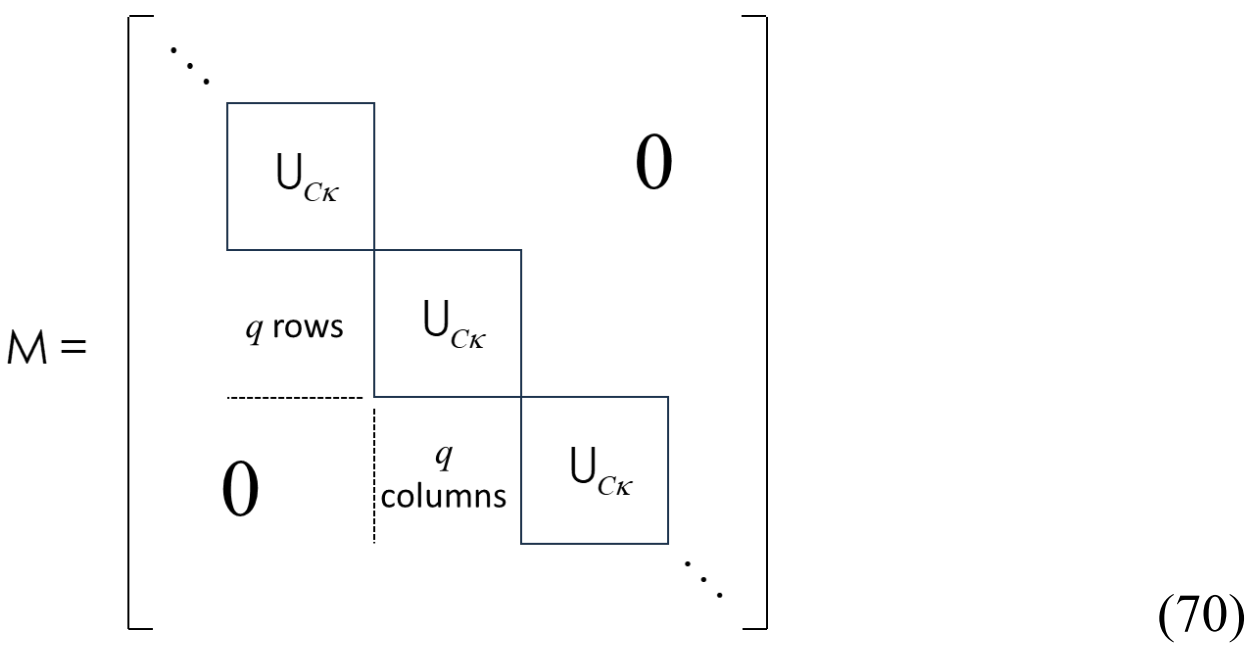

(70)

The problem of evaluating eigenvectors and eigenvalues is now reduced just to solving for the eigenvectors and eigenvalues of a square matrix $\mathsf{U}_{C\kappa}$ of size $2m$, where $m$ is the number of rows of blocks in the periodic mesh. This will enable us to calculate a "band structure" of phase shifts $\eta$ as a function of the "wavevector" $\kappa$. Note that, if the physical inputs to the mesh are equally spaced by some amount $a$, we can evaluate the corresponding physical wavevector component $k$ along the input surface of the mesh by multiplying by $a$, i.e.,

$$k = \kappa a \tag{71}$$

hence establishing physical eigenvectors for the incoming waves.

The self-configuring algorithm for setting up the mesh in this periodic case is given in detail below in Supplemental document section S8. For the simpler special case where the configuration Vectors are all just correspondingly shifted versions of Vector 1, which we could refer to as simple periodicity, the resulting algorithm is given below in Supplemental document section S9. The spatial filter based on Daubechies wavelets discussed in the main text actually corresponds to this simpler case.

The eigenfunctions for this forward photonic crystal are the transverse functions that reproduce themselves with a phase factor between the input – the "top" of the multilayer structure – and the output – the "bottom" of the multilayer structure as in Fig. S4. (The output functions may be moved slightly sideways compared to the input functions, by $m-1$ "positions" if we draw the mesh with rectangular blocks and vertical connecting lines or "waveguides" between the blocks as in Fig. 4 of the main text, but if we use the numbering system for inputs and outputs as in Fig. 4 of the main text, the eigenfunctions reproduce themselves between inputs and outputs.)

It would also be possible to stack multiple such sets of $m$ layers one on top of the other as sketched in Fig. S5, showing three such sets of layers.

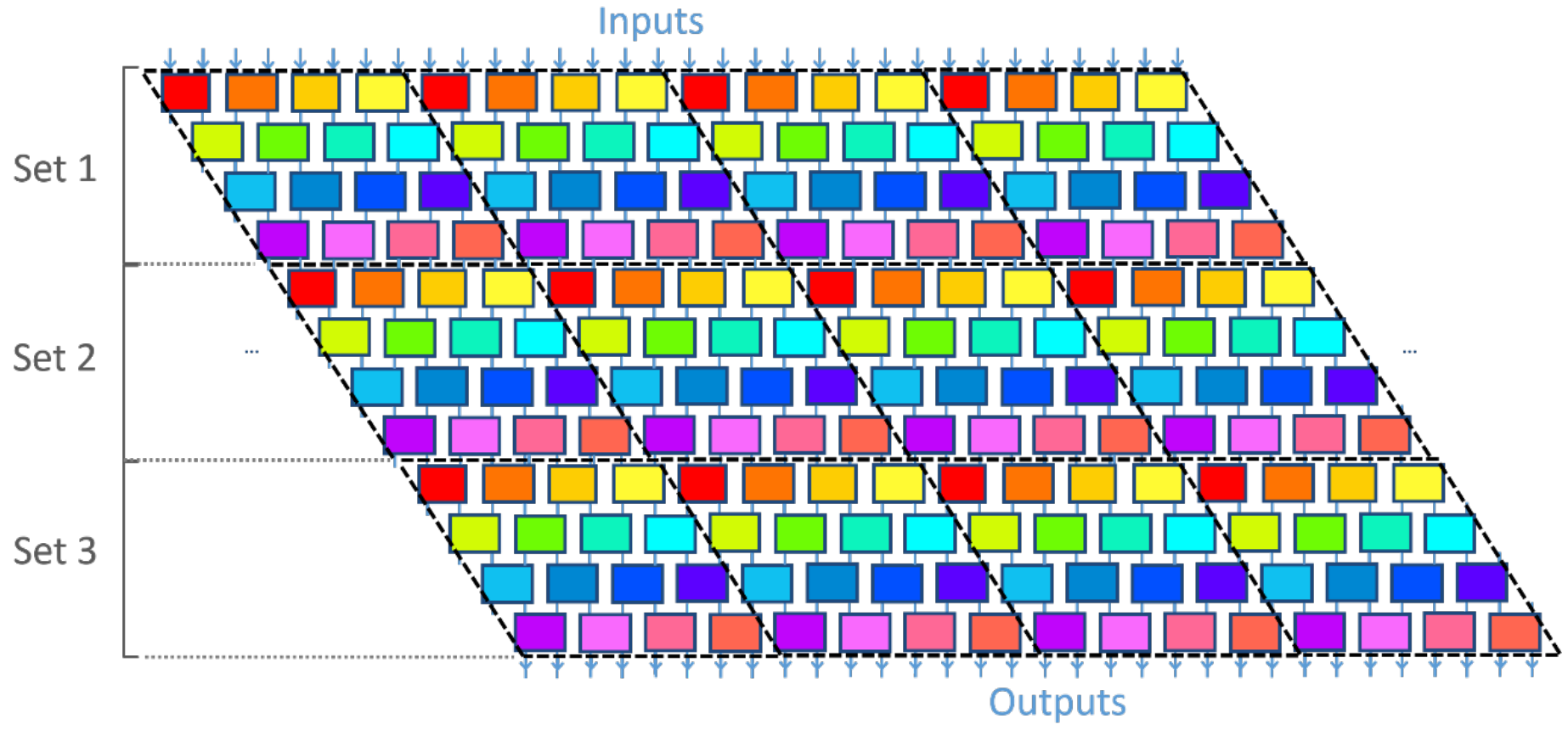

Fig. S5. Stacking of multiple sets of m layers of blocks, forming a "forward photonic crystal" also now in the vertical direction.

In this case the structure would also become periodic in the "vertical" direction (or, more precisely, along the direction of the sides of the parallelogram unit cells), and the eigenfunctions introduced here would reproduce themselves in each successive set of $m$ layers, with a phase factor of $\exp(i\eta_\kappa)$ between successive sets of $m$ layers. So, if we index the successive sets of $m$ layers of blocks by index $Q$, we have

$$\left|\psi_{Cp\kappa,Q+1}\right\rangle = \exp\left(i\eta_{p\kappa}\right)\left|\psi_{Cp\kappa,Q}\right\rangle \tag{72}$$

which is a Bloch-like form as in Eq. (34), but now in this "vertical" direction. In this sense, we would now have a 2-dimensional photonic crystal structure.

The effect of using $n$ such sets of layers in sequence would be simple to magnify the eigenvalue phase shifts by $n$ – i.e.,

$$\left|\psi_{Cp\kappa,Q+n}\right\rangle = \exp\left(in\eta_{p\kappa}\right)\left|\psi_{Cp\kappa,Q}\right\rangle \tag{73}$$

Note that this is a kind of "logical" crystal structure in the vertical direction, being periodic in the sets of blocks. Those sets of blocks would not have to be precisely spaced vertically to get this kind of Bloch-like behavior with respect to the index $p$, at least for the relative phase between eigenfunctions of different $\kappa$ and $p$. Such overall spacing of these sets of layers would affect the absolute phase of light propagating through the structure, but not the relative phases of the different eigenstates.

### *Non-unitary periodic structure*

We could use the non-unitary construction as in Fig. 9 in the main text in a lateral forward photonic crystal configuration, following on from the unitary case of Fig. S4. We would use a parallelogram of blocks to feed through the amplitude modulators and phase shifters to another parallelogram of blocks, as in the example unit cell in Fig. S6.

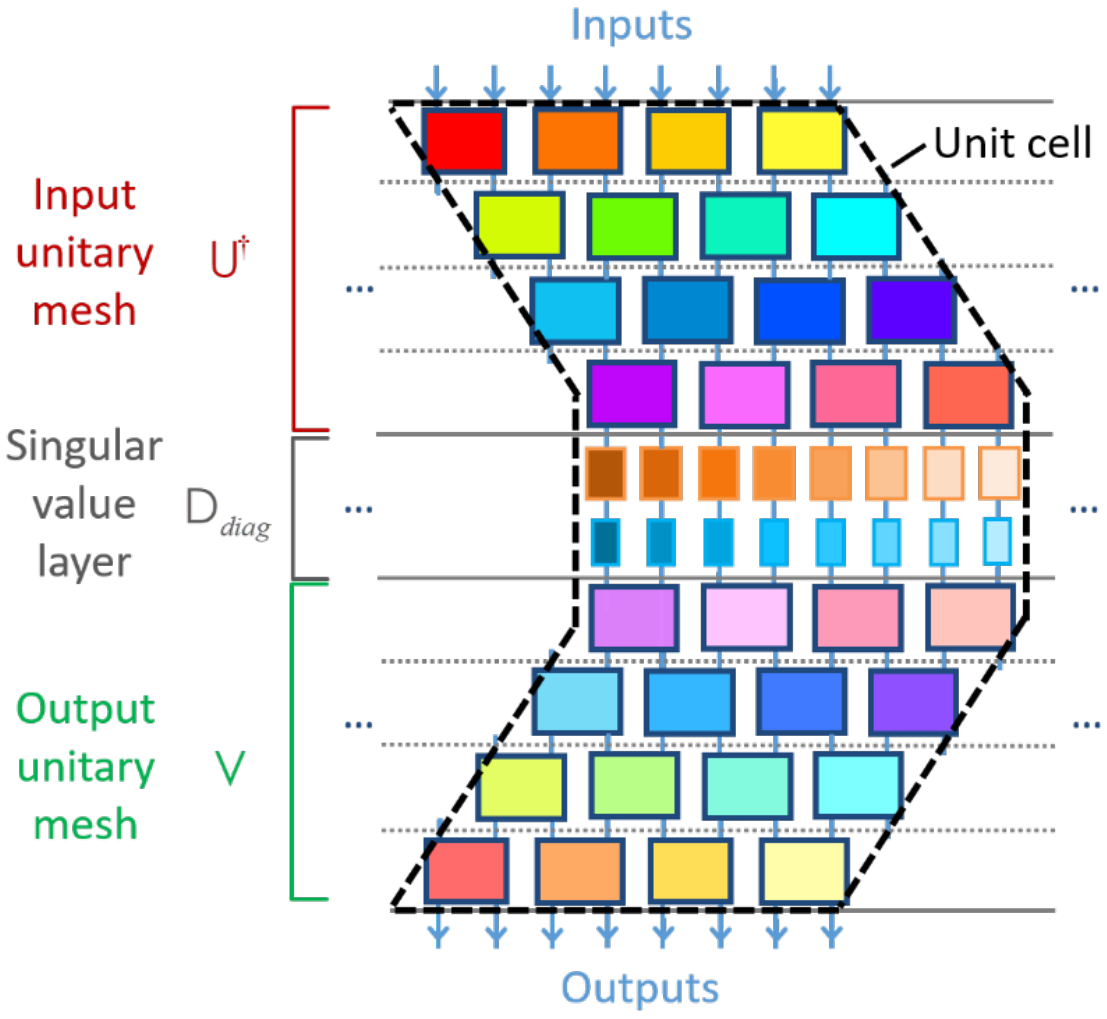

Fig. S6. Example unit cell for a non-unitary forward photonic crystal mesh, here for 4 layers in each unitary mesh. In this example the output unitary mesh is oriented in the opposite direction to the input unitary mesh to realign the Inputs and Outputs.

## S8 Self-configuring algorithm for periodicity of *m* blocks for an *m* block layer mesh

If the mesh is to be space-invariant in groups of size equal to the number of layers of $2\times 2$ blocks (here 4 layers), so block B$i(j+4)$ is set to be the same as block B$ij$ (e.g., B25 is set the same as B21), then we can use the following algorithm. This algorithm presumes that the blocks are calibrated or identically fabricated so we can replicate the settings from blocks in one unit cell to those in the next. If this is not the case, then we can configure with the algorithm in Supplemental document Section S5 above using shifted versions of the configuration Vectors for successive unit cells or diagonal lines.

Layer 1 settings
- Shine in only Vector 1
  - Adjust block B11 to set the output power in its left output port to zero
    - Replicate settings to B15
  - Adjust block B14 to set the output power in its right output port to zero.
- Comment: these steps ensure both that power is flowing "rightwards" from B11 towards Out 1 and "leftwards" from B14 towards Out 1
- Shine in only Vector 2
  - Adjust block B12 to set the output power in its left output port to zero
    - Replicate settings to B16
      - Comment: We could then adjust block B15 to set the output power in its right output port to zero, but B15 will already have been set as the replica of B11.
- Shine in only Vector 3
  - Adjust block B13 to set the output power in its left output port to zero
    - Replicate settings to B17
      - Comment: We could then adjust block B16 to set the output power in its right output port to zero, but B16 will already have been set as the replica of B12.
- Comment: We do not need to shine in Vector 4 because there is nothing left for it to set in this layer.

Layer 2 settings
- Shine in only Vector 1

Adjust block B21 to set the output power in its left output port to zero
Replicate settings to B25
Adjust block B23 to set the output power in its right output port to zero.
Comment: these steps ensure both that power is flowing "rightwards" from B21 towards Out 1 and "leftwards" from B23 towards Out 1
Shine in only Vector 2
Adjust block B22 to set the output power in its left output port to zero
Replicate settings to B26
Adjust block B24 to set the output power in its right output port to zero
Layer 3 settings
Shine in only Vector 1
Adjust block B31 to set the output power in its left output port to zero
Replicate settings to B35
Adjust block B32 to set the output power in its right output port to zero.
Shine in only Vector 2
Adjust block B33 to set the output power in its right output port to zero
Shine in only Vector 3
Adjust block B34 to set the power in its right output port to zero
Layer 4 settings
Shine in only Vector 1
Adjust block B41 to set the power in its right output port to zero
Shine in only Vector 2
Adjust block B42 to set the power in its right output port to zero
Shine in only Vector 3
Adjust block B43 to set the power in its right output port to zero
Shine in only Vector 4
Adjust block B44 to set the power in its right output port to zero

This process will have set all the blocks B11 to B44. For this horizontally periodic system, we could then proceed to replicate these blocks periodically to the right, stepping by 4 blocks to the right to get a periodic structure with period 4 blocks in the "horizontal" direction. Note that the blocks B15, B16, B17, B25, B26 and B35 have already been set this way by the replications mentioned in the algorithm above.

### S9 Self-configuring algorithm for setting the block layer mesh for simple shifted inputs

This algorithm can configure a mesh that has simple periodicity – that is, every block in a given layer is set the same so the mesh is space-invariant as we move by any integer number of blocks.

Comment: perform "collection triangle" settings for Output 1
Layer 1 settings
Shine in only Vector 1
Adjust B11 to set the output power in its left output port to zero
Comment: This is required because we need all the power to flow towards the bottom block B14, which it can only do if B11 routes its output power to its right port.
Comment: At this point we could adjust B14 to set the output power in its right output port to zero; this is required because we need all the power to flow towards the bottom block B14, which it can only do if B14 routes its output power to its right port. However, for simplicity of the algorithm we do not need to do this here because we can set this same block later using Vector 4. The fact that there are two algorithmic ways of setting this block is actually telling us there is a necessary constraint on the last two elements of Vector 1, which is that that two element vector is orthogonal to the two element vector that is the first two elements of Vector 1

Shine in only Vector 2 (which is just a shifted version of Vector 1 in this case)

Adjust B12 to set the output power in its left output port to zero

Comment: This is required because we need all the power to flow towards the bottom block B14, which it can only do if B11 routes its output power to its right port.

Shine in only Vector 3 (which is just a shifted version of Vector 1 in this case)

Adjust B13 to set the output power in its left output port to zero

Shine in only Vector 4 (which is just a shifted version of Vector 1 in this case)

Adjust B14 to set the output power in its left output port to zero

Layer 2 settings

Shine in only Vector 1

Adjust B21 to set the output power in its left output port to zero

Shine in only Vector 2

Adjust B22 to set the output power in its left output port to zero

Shine in only Vector 3

Adjust B23 to set the output power in its left output port to zero

Layer 3 settings

Shine in only Vector 1

Adjust B31 to set the output power in its left output port to zero

Shine in only Vector 2

Adjust B32 to set the output power in its right output port to zero

Layer 4 settings

Shine in only Vector 1

Adjust B41 to set the output power in its right output port to zero

Comment: this completes the setting if the "collection triangle" of blocks for Vector 1 to be routed to Output 1. We can now proceed to set the remaining blocks, one successive diagonal line at a time with one successive input Vector.

Comment: successive diagonal line settings

Shine in only Vector 2

Adjust B15 to set the output power in its right output port to zero
Adjust B24 to set the output power in its right output port to zero
Adjust B33 to set the output power in its right output port to zero
Adjust B42 to set the output power in its right output port to zero

Shine in only Vector 3

Adjust B16 to set the output power in its right output port to zero
Adjust B25 to set the output power in its right output port to zero
Adjust B34 to set the output power in its right output port to zero
Adjust B43 to set the output power in its right output port to zero and so on for successive (shifted) input vectors and additional corresponding diagonal lines of blocks on the right in Fig. 4 of the main text.

If the blocks are physically calibrated, then we can simply replicate the settings of the first block in a block layer to all subsequent blocks in that layer.